\documentclass[aps,physrev,groupedaddress,twocolumn]{revtex4-2}
\usepackage{graphicx}
\usepackage{dcolumn}
\usepackage{bm}
\usepackage{multirow}
\usepackage{graphicx}
\usepackage{amsmath} 
\usepackage{hyperref}
\usepackage{gensymb}
\usepackage{float}
\usepackage{xcolor}

\begin{document}
\title{Particle Acceleration in Cassiopeia A Revealed by Broadband High-Energy Spectrum}

\author{Bo-Tao Li}
\affiliation{Department of Astronomy, School of Physics and Technology, Wuhan University, Wuhan 430072, China}

\author{Wei Wang}
\altaffiliation{Email address: wangwei2017@whu.edu.cn} 
\affiliation{Department of Astronomy, School of Physics and Technology, Wuhan University, Wuhan 430072, China}

\author{Zhuo Li}
\altaffiliation{Email address: zhuo.li@pku.edu.cn}
\affiliation{Department of Astronomy, School of Physics, Peking University, Beijing 100871, China}
\affiliation{Kavli Institute for Astronomy and Astrophysics, Peking University, Beijing 100871, China}

\date{\today}

\begin{abstract}
Recently, the GeV--sub-PeV spectrum of supernova remnant (SNR) Cassiopeia A (Cas A), one of the youngest and most well-studied SNRs in our Galaxy, has been updated by observations of Fermi-LAT and LHAASO. We revisit Cas A with our previous shell-plus-jet asymmetric model and investigate its particle acceleration ability. The broadband fitting results suggest that the double-peaked gamma-ray spectrum can be well attributed to proton-proton (PP) collisions and inverse Compton scattering within the SNR shell, while the synchrotron emission from a jet component with velocity of $\sim0.1c$ can account for the hard X-ray emission up to 220 keV. Furthermore, the PP collisions in the jet can produce a sub-PeV emission, but constrained by the LHAASO-KM2A limit to a flux below $\sim 1\times10^{-14}\rm erg/(cm^2s)$ at 100 TeV. The energy of accelerated protons in the jet of Cas A could be up to $5\times10^{47}$ erg, which, assuming that the PeV cosmic ray distribution is clumpy in the Galaxy with the clump size comparable to the thickness of the Galactic plane, derives a proton flux consistent with the observed one at 1 PeV, implying that the Cas A-like SNRs can still be PeVatrons in the Galaxy.
It is encouraging for LHAASO and future telescopes to detect or constrain Cas A spectrum above 100-TeV more precisely. 
\end{abstract}


\maketitle

\section{Introduction}
\label{sec:intro}

Supernova remnants (SNRs) are extended structures resulting from the explosive death of stars and have long been considered to be the primary candidate for the acceleration of cosmic rays (CRs) \citep[e.g.,][]{2012APh....39...52D}. These explosions under extreme conditions will eject material at high velocities and generate powerful shock waves that heat up surrounding medium and accelerate charged particles, say, electrons and protons. The acceleration mechanism is widely believed to be diffusive shock acceleration (DSA), in which particles repeatedly cross the shock front, gaining energy in each cycle \citep{krymskii1977regular,bell1978acceleration,1978ApJ...221L..29B,drury1983introduction,1987PhR...154....1B}. Studying the non-thermal emission radiated by accelerated particles will provide insights into particle energy distribution and acceleration mechanism within SNRs. Non-thermal radiation processes in the shock region include synchrotron radiation (Syn), inverse Compton scattering (IC), non-thermal Bremsstrahlung (Bre), and decay of neutral pions resulting from proton-proton collisions (PP). 

Cassiopeia A (Cas A) is a SNR located approximately \(3.4^{+0.3}_{-0.1}\) kpc away from Earth \citep{reed1995three}. Believed to have exploded between the 1670s and 1680s, Cas A is one of the youngest known SNRs in the Galaxy. Optical spectral analyses have identified that it resulted from a type IIb supernova \citep{krause2008cassiopeia}. Cas A is very bright in multi-bands, thus it has been intensely observed by many telescopes covering a broad energy range, making it an excellent subject for studying particle acceleration in SNRs.

Radio \citep[e.g.,][]{bell1975new,delaney2014density} and X-ray mapping \citep[e.g.,][]{hwang2004million} revealed the structure of Cas A as an expanding shell with a bright ring and a faint ring, with radii of $\sim\mathrm{1.7^\prime/1.7\ pc}$ and $\sim\mathrm{2.5^\prime/2.5\ pc}$, respectively, assuming a distance of 3.4 kpc, which are commonly interpreted as regions of reverse and forward shock. The shock radius ratio may indicate that Cas A's shock has just transitioned from free expansion to Sedov phase.

The broadband SED spectrum of Cas A shows two spectral bumps. The low-energy bump from radio to X-ray band is generally believed to be Syn dominated, but whether the gamma-ray bump results from lepton or baryon processes is still under extensive discussion \citep[e.g.,][]{2010ApJ...720...20A,2014ApJ...785..130Z,zhang2019supernova,zhan2022asymmetrical}. Notably, many works have discussed the insufficiency of a single-zone model to explain the SED of Cas A. \citet{atoyan2000energy}, based on radio imaging and spectral data, proposed a two-zone model for the spatially non-uniform shell, which distinguished the radio knots/rings and the rest diffuse region with high and low densities of accelerated electrons, respectively.

The INTEGRAL-IBIS data showed that the hard X-ray spectrum follows a power-law without a spectral cutoff up to 220 keV \citep{wang2016hard}. The absence of spectral cutoff is inconsistent with the expected cutoff at $\sim3.8\eta_{\rm g}^{-1}(v_s/5000\, {\rm km\,s^{-1}})^2$ keV (with $\eta_{\rm g}>1$ the particle diffusion coefficient in unit of that of the Bohm limit, and $v_s$ the forward shock velocity) from the DSA process. This paper argues against the origin of the hard X-ray emission from the central compact object, the magnetic field inhomogeneity, and the PP-induced secondary emission, etc, but favors a Syn origin from a jet component with high velocity of $\sim0.1c$. Additionally, the estimation of \(^{44}\mathrm{Ti}\) yields \((1.3\pm0.4)\times10^{-4}M_{\odot}\), which also supports the hypothesis of an asymmetric and/or a more energetic explosion \citep{wang2016hard}. 

Assuming a distance of 3.4 kpc, X-ray imaging \citep{hwang2004million} and optical study on high speed knots \citep{fesen2006expansion} both suggest that Cas A consists of a primary shell and a bipolar jet moving at transverse velocities of $\sim$ \(5000\mathrm{km\ s^{-1}}\) and $\sim$ \(14000 \mathrm{km\ s^{-1}}\), respectively. According to \citet{wang2016hard}, the relation between velocity and opening angle is given by $v\simeq v_0(\theta/\theta_{0})^{-0.27}$. Considering the shell component with $\theta_0\sim\pi$ and $v_0\sim5000\, \rm km \, s^{-1}$, a jet velocity with $v\gtrsim 10^4\,\rm km \, s^{-1}$ corresponds to an opening angle of $\theta\lesssim10^{\circ}$, consistent with observed $\theta_{\rm jet}\sim12.5^{\circ}$ \citep{fesen2006expansion}. Considering projection effect, the true velocity of jet component could be up to $\sim0.1c$. 

For the broadband spectrum modeling, \cite{zhang2019supernova}, following the reasoning of \citet{atoyan2000energy}, applied a two zone model, but explained the two zones as forward and reverse shocks. On the other hand, \cite{zhan2022asymmetrical} applied a shell-plus-jet asymmetrical model following \cite{wang2016hard}, and explained the high energy gamma-ray bump as contributions from PP and IC from the SNR shell, and predicted a sub-PeV emission from the jet component.  

Recently, the LHAASO collaboration published new gamma-ray results from LHAASO-KM2A \citep{cao2024does} and LHAASO-WCDA \citep{cao2025broadband}, as well as the updated Fermi-LAT result. In this study, we adopt the asymmetrical model to reinvestigate the broadband spectrum of Cas A and study its particle acceleration ability under the framework of DSA theory. The paper is organized as follows. Dataset used is introduced in section ~\ref{sec:data}. Details of the two-zone emission model are presented in section ~\ref{sec:model}. The fitting results are shown in section ~\ref{sec:res}, followed by discussion and comparison with previous work in section ~\ref{sec:dis}. Finally, we draw the conclusion in section ~\ref{sec:con}. 

\begin{figure}
    \centering
    \includegraphics[width=1\linewidth,height=1\linewidth]{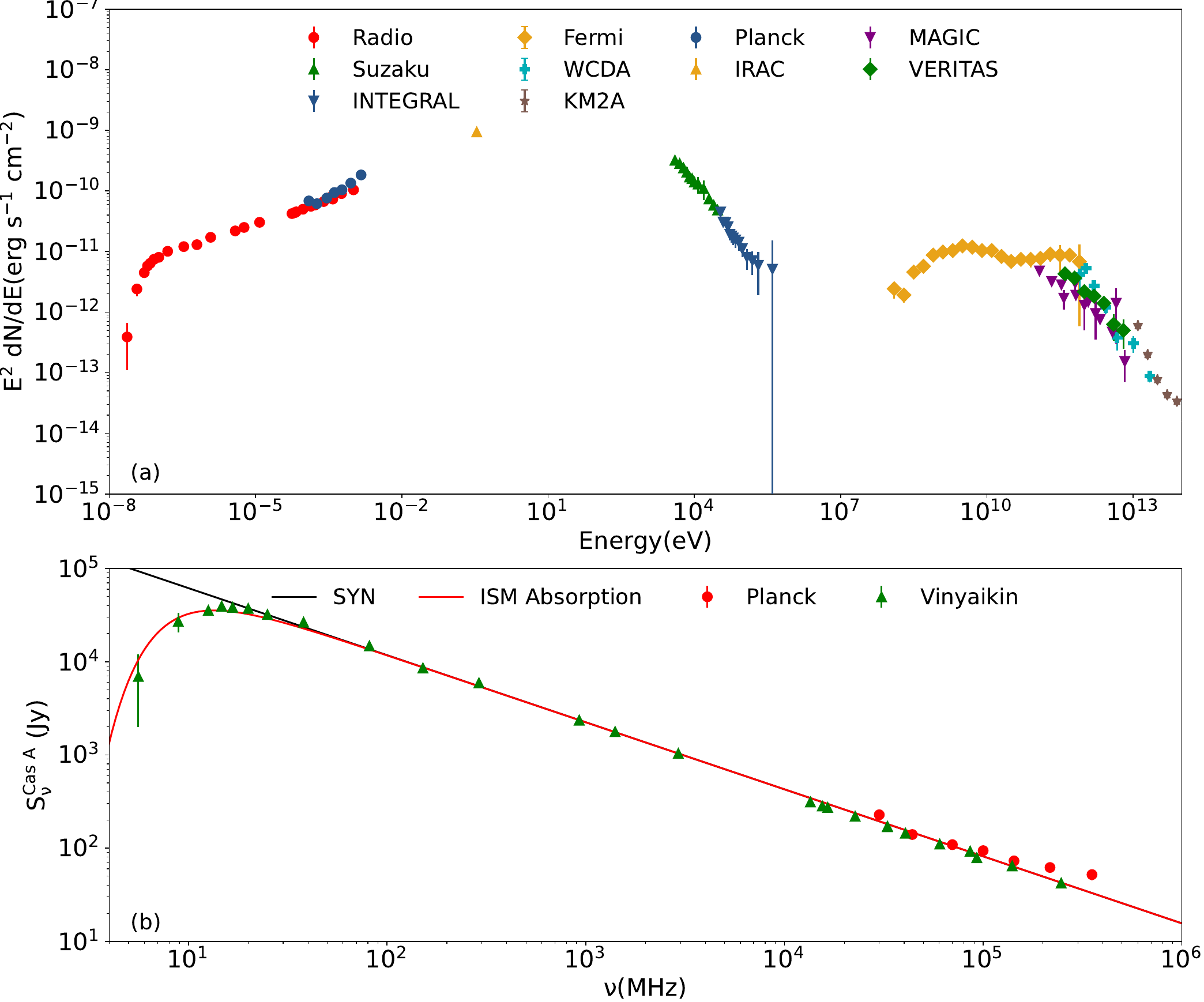}
    \caption{Upper panel: The broadband spectrum of Cas A, $E^2dN/dE$ versus energy. Observational data are taken from \citet{vinyaikin2014frequency} (radio), \citet{arnaud2016planck} (Planck), \citet{de2016dust} (IRAC), \citet{maeda2009suzaku} (Suzaku), \citet{wang2016hard} (INTEGRAL-IBIS), \citet{ahnen2017cut} (MAGIC),  \citet{abeysekara2020evidence,humensky2008veritas} (VERITAS), and \citet{cao2025broadband} (Fermi-LAT and LHAASO). Lower panel: radio spectrum, flux density versus frequency, fitted by power law with free-free absorption. }
    \label{fig:spec}
\end{figure}

\section{Broadband spectrum data}
\label{sec:data}

Cas A is the brightest extrasolar radio source in the sky at frequencies above 1 GHz, with a flux density of \(S_{1\mathrm{GHz}}=2723\mathrm{\ Jy}\) in 1980, and has been decaying at a rate of \(\mathrm{0.97(\pm0.04)-0.30(\pm0.04)\log(\nu/GHz) \%/yr}\) \citep{baars1977absolute}.

Cas A is also one of the brightest X-ray emitters in the sky. In the 3.4-40 keV band, the Suzaku spectrum is best described as a thermal bremsstrahlung plus a non-thermal cutoff power law with an index of $\Gamma=2.2 \pm 0.7$ and a cutoff at $3.4^{+\infty}_{-1.7}$ keV \citep{maeda2009suzaku}. RXTE data in the 2-60 keV band show a broken power law of $\Gamma_1=1.8^{+0.5}_{-0.6}$ and $\Gamma_2=3.04^{+0.15}_{-0.13}$ below and above energy \(E_c=15.9^{+0.3}_{-0.4}\ \mathrm{keV}\) \citep{allen1997evidence}. INTEGRAL-IBIS observations yield similar results: the emission from 3 to 500 keV can be fitted with a thermal bremsstrahlung component plus a power law of $\Gamma= 3.13\pm0.03$. Notably, up to 220 keV, the IBIS spectrum does not show a spectral cutoff \citep{wang2016hard}.

Due to the limited spatial resolution, gamma-ray observations can only treat Cas A as a point source. HEGRA first detected emission between 1-10 TeV with a power-law spectrum of $\Gamma=2.5\pm0.4_{\rm stat}\pm0.1_{\rm sys}$ \citep{aharonian2001evidence}. Subsequent MAGIC observations measured a power-law spectrum with index $\Gamma=2.3\pm0.2_{\rm stat}\pm0.2_{\rm sys}$ above 250 GeV \citep{albert2007observation}. Fermi-LAT filled the observation gap between 0.5 GeV and 50 GeV, reporting a power-law spectrum with index $\Gamma=2.0\pm0.1$ without significant exponential cutoff \citep{abdo2010fermi}. Furthermore, MAGIC observations indicated that the spectrum is best fitted by an exponential cutoff power law (ECPL) with $\Gamma=2.4\pm0.1_{\rm stat}\pm0.2_{\rm sys}$ and an exponential cutoff at \(E_c=3.5(^{+1.6}_{-1.0})_{\rm stat}(^{+0.8}_{-0.9})_{\rm sys}\) TeV \citep{ahnen2017cut}. The combination of observations by Fermi-LAT and MAGIC implies that the emission from GeV to 10 TeV may be dominated by a single-peaked component around 10 GeV \citep{ahnen2017cut}.

Recently, LHAASO-WCDA observations combined with updated Fermi-LAT data have revealed new features in the high energy spectrum \citep{cao2025broadband}. The spectrum shows a double-bump structure peaking at around GeV and TeV, respectively, with a dip at about 20 GeV; beyond 20 GeV the spectrum is a broken power law with a break energy of \(0.63\pm0.21\) TeV. The spectrum from 1 TeV to 10 TeV by LHAASO-WCDA is steeper but higher than that reported by imaging air Cherenkov telescopes (IACTs), MAGIC and VERITAS. At higher energy, i.e., 10 TeV to 1 PeV, LHAASO-KM2A observation has also placed upper limit on the flux from Cas A \citep{cao2024does}. 

Spectral fitting in this study utilizes radio data from \citet{vinyaikin2014frequency}, Suzaku data from \citet{maeda2009suzaku}, INTEGRAL-IBIS data from \citet{wang2016hard}, and Fermi-LAT and LHAASO data from \citet{cao2025broadband,cao2024does}, as shown in the upper panel of Figure ~\ref{fig:spec}. 

It should be noted that radio band and infrared band can be strongly affected by absorption and continuum radiation from surrounding dust. To model the Syn component, we compare radio dataset of \citet{vinyaikin2014frequency} with Hershel infrared data and Planck microwave survey of \citet{de2016dust,arnaud2016planck} in the upper panel of Figure ~\ref{fig:spec}. In \citet{de2016dust}, IRAC $\mathrm{3.6\ \mu m}$ data are considered to be dominant by Syn and used to model Syn emission together with Planck data at 44, 70, 100 and 143 GHz. Considering that infrared band can be easily affected by processes such as thermal emission, and that \citet{vinyaikin2014frequency} provides a dataset with longer time spans, we adopt the radio dataset from \citet{vinyaikin2014frequency} in our spectrum fitting. 

Moreover, located in the Galactic plane, Cas A is inevitably affected by severe free-free absorption by external interstellar medium (ISM) at low-frequency radio band \citep{de2020cassiopeia,stanislavsky2023free,arias2018low}. With $Z$, $T_e$ and EM being the average number of ion charge, the electron temperature of the absorbing medium and emission measure, we can evaluate the radio flux density at frequency $\nu$ with free-free attenuation as $S_{\rm obs}(\nu)=S(\nu)e^{-\tau_{\rm ISM}(\nu)}$ \citep{stanislavsky2023free}, where the free-free optical depth of ISM $\tau_{\rm ISM}$ is:
\begin{align}
\tau_{\rm ISM}(\nu) = & 3.014 \times 10^4
\left( \frac{\nu}{\mathrm{MHz}} \right)^{-2}
\left( \frac{T_e}{\mathrm{K}} \right)^{-3/2}
 \left( \frac{\mathrm{EM}}{\mathrm{pc\;cm^{-6}}}\right) \notag \\
  &\times Z\ \ln \left[ 
\frac{49.55}{Z}
\left( \frac{T_e}{\mathrm{K}} \right)^{3/2} 
\left( \frac{\nu}{\mathrm{MHz}} \right)^{-1}
\right].
\label{eq:tau}
\end{align}
Here we approximate the radio low-frequency turnover as free-free absorption in ISM using $Z=1$, $T_e=50K$ and $\rm EM=0.1\ pc\ cm^{-6}$. The absorption effect of low-frequency radio data is shown in the lower panel of Figure ~\ref{fig:spec}.  

\section{Radiation model}
\label{sec:model}

Following \citet{zhan2022asymmetrical}, we apply an asymmetric ejecta model simplified as a primary SNR shell and a bipolar jet, with velocities of $v_{\rm shell}=5000~\mathrm{km\, s^{-1}}$ and $v_{\rm jet}=30000~\mathrm{km\, s^{-1}}$, respectively, corresponding to a velocity ratio $v_{\rm jet}/v_{\rm shell}=6$ and an angle of approximately $30^\circ$ between the jet axis and the line of sight. 

\begin{figure}
    \centering
    \includegraphics[width=1\linewidth]{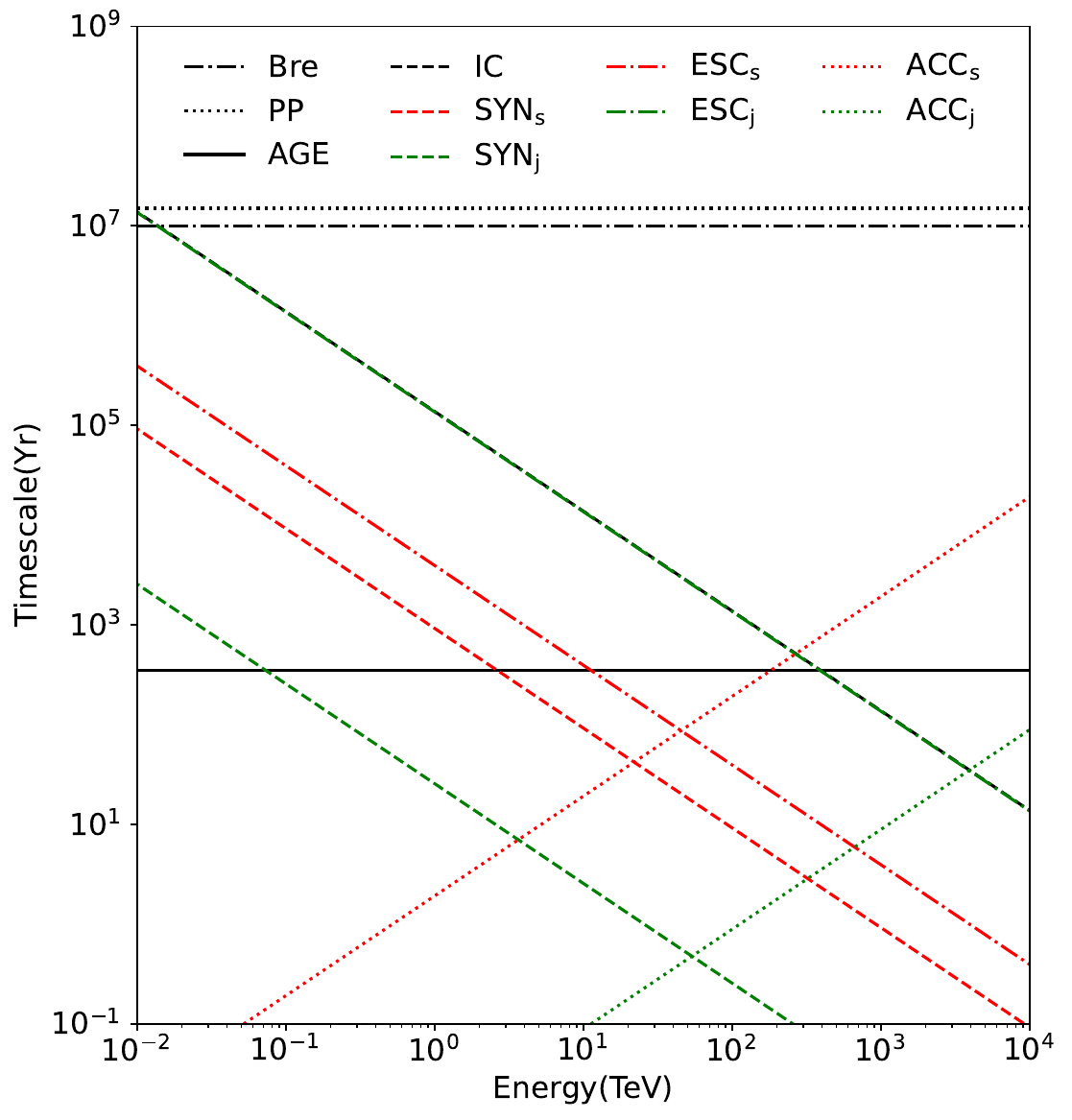}
    \caption{Timescales related to particle acceleration and radiation processes as a function of particle energy in model M3. The red and green lines represent timescales for the shell and jet zones, respectively, while black lines represent those shared timescales. }
    \label{fig:timescale}
\end{figure}

The shell and jet interacting with medium generate shocks in the edges. The shocks are expected to accelerate charged particles and amplify magnetic field in the swept-up medium. We follow the physical model of DSA mechanism to consider the particle acceleration and the particle energy distribution in the shocks. The DSA mechanism generally accelerates charged particles to higher energy following a power-law distribution, $dN/dp\equiv N(p)\propto p^{-\alpha}$, with $p$ being the particle momentum \citep{bell1978acceleration,1978ApJ...221L..29B,1987PhR...154....1B}. In non-relativistic (NR) strong shocks, under test particle approximation, the particle index is exactly $\alpha=2$ \citep[e.g.,][]{1987PhR...154....1B}, whereas the universal prediction for relativistic shock is $\alpha=2.23$ \citep{2000ApJ...542..235K,2005PhRvL..94k1102K}. 

Regarding radiation by particles, we consider relativistic particles and neglect NR ones so that we have particle energy distribution $N(E)\propto E^{-\alpha}$, since $E\propto p$ for relativistic particles. However, a turnover appears around $E\sim 2mc^2$ in the energy distribution, with $m$ being the particle rest mass, and the total energy is dominated by relativistic particles with $E\gtrsim2mc^2$, if $\alpha\leq3$ \citep{2013ApJ...778..107S}. Thus for NR shock, we set a lower limit on the accelerated protons that 
\begin{equation}
    E_{\min,p}\simeq 2m_pc^2.
\end{equation} 
For accelerated electrons, if their energy shares a fraction $\epsilon_e$ of the shock energy, the lower limit can be set as equation (6) in \cite{2018ApJ...859..123H}, 
\begin{equation}
    E_{\min, e}=\max\{(\alpha-2)(\alpha-1)^{-1}m_p\epsilon_ev^2,\ 2m_ec^2\},
\end{equation}
where $v$ is the ejecta velocity, i.e, $v_{\rm shell}$ or $v_{\rm jet}$. We take the typical value $\epsilon_e=0.1$. 

As the DSA mechanism does not distinguish electrons and protons, we should assume that electrons and protons follow the same index $\alpha$. The power law $E^{-\alpha}$ only extends up to some finite energy, which is then limited by the particle acceleration ability of the shock. Moreover, the particle energy distribution in the shocks can be affected by radiation losses, thus deviates from the accelerated power law. Considering all these effects, the energy distributions of relativistic electrons and protons in the shocks follow, respectively,
\begin{align}
N_e(E) &= A_e \left( \frac{E}{1\,\mathrm{TeV}} \right)^{-\alpha} \exp\left( -\frac{E}{E_{\mathrm{cut,e}}} \right)\notag\\ &\times
\begin{cases}
1 & E_{\min,e}< E \leq E_b \\
(E/E_b)^{-1} & E > E_b
\end{cases},
\label{eq:ecbpl}\\
N_p(E) &= A_p \left(\frac{E}{1\,\mathrm{TeV}}\right)^{-\alpha} \exp\left(-\frac{E}{E_{\rm cut,p}}\right) & E>E_{\min,p}.
\label{eq:ecpl}
\end{align}
Given the distributions above, the total energies of relativistic electrons and protons can be calculated as $W_e=\int_{E_{\min,e}} EN_e(E)dE$ and $W_p=\int_{E_{\min,p}} EN_p(E)dE$, respectively. The lower limits take the low energy limits by relativistic particles condition into account. 

There are two more characteristic energies in the particle energy distribution. Cutoff energy $E_{\rm cut, e/p}$ is the maximum energy that particles can be accelerated to, and is determined by the balance between particle acceleration timescale $t_{\rm acc}$ and cooling timescale $t_{\rm loss}$, including Syn cooling $\tau_{\rm syn}$, IC cooling $\tau_{\rm IC}$, Bre cooling $\tau_{\rm bre}$, PP energy loss $\tau_{\rm pp}$ and timescale $\tau_{\rm esc}$ of particles' escaping upstream from free escape boundary \citep[e.g.,][]{2008ApJ...678..939Z,2011ApJ...731...87E}, and the limitation of the SNR age $t_{\rm age}\approx350$ yr. We can write the balancing equation as $1/t_{\rm acc}=\max\{1/t_{\rm loss,},\ 1/t_{\rm age},1/\tau_{esc}\}$. As the dominant cooling processes for electrons and protons are different, we have
\begin{align}
\frac{1}{t_{loss}}=
\begin{cases}
\frac{1}{\tau_{IC}}+\frac{1}{\tau_{syn}}+\frac{1}{\tau_{bre}} & \mathrm{electrons}, \\
\frac{1}{\tau_{pp}} & \mathrm{protons}.
\end{cases}
\label{eq:cutoff}
\end{align}
The calculation of all the related timescales are presented in appendix \ref{app:timescale}. Figure ~\ref{fig:timescale} illustrates, for example, the timescales in the case of model M3 (see Section~\ref{sec:res} and Table \ref{tab:results}), where the maximum energies of electrons and protons are mainly limited by Syn cooling ($t_{acc}=\tau_{syn}$) and particle escaping upstream ($t_{acc}=\tau_{esc}$), respectively, i.e.,
\begin{align}
E_{\rm cut,e}\simeq\frac{40}{(\eta_gm)^{1/2}}\left(\frac{B_{d}}{\rm 100\mu G}\right)^{-1/2}\left(\frac{v_s}{5\times10^8\rm cm\,s^{-1}}\right)\rm TeV,
\label{Max_esc}
\end{align}
\begin{align}
E_{\rm cut,p}\simeq\frac{11.3}{\eta_gm}\left(\frac{B_d}{\rm 600\mu G}\right)\left(\frac{v_s}{0.1c}\right)\left(\frac{\kappa R}{\rm 0.3pc}\right)\rm PeV.
\label{Max_syn}
\end{align}
Definitions of parameters $\eta_g$, $m$ and $\kappa$ are given in appendix \ref{app:timescale}.

In addition to maximum energy limitation, radiative cooling will affect the energy distribution by introducing a break energy $E_b$, for electrons mainly through synchrotron (SYN) and inverse Compton (IC) processes. For Cas A, under reasonable parameter conditions, the cooling is constrained by synchrotron radiation, as illustrated in Figure ~\ref{fig:timescale}. High energy electrons cool down very fast within a dynamical time, i.e., the SNR age, causing a spectral steepening from index $\alpha$ to $\alpha+1$ above $E_b$, which is determined by equating Syn cooling timescale and the SNR age, $\tau_{syn}=t_{age}$,
\begin{align}
E_b\simeq4\left(\frac{B_{d}}{100\rm \mu G}\right)^{-2}\left(\frac{t_{age}}{350\rm yr}\right)^{-1}\rm TeV.
\label{eq:broken}
\end{align}

Provided the particle distribution, the calculation of their radiation still needs the background physical conditions, described below. First, for PP and Bre processes one needs the number densities of protons and ions. As X-ray spectra of Cas A indicate an hydrogen column density corresponding to a progenitor star wind density of $0.9\pm0.3\ cm^{-3}$ \citep{lee2014x}, we assume a downstream density in both the SNR shell and jet of $n_H=3.6\ \rm  cm^{-3}$, as the medium density is compressed by a factor of 4 in a NR, strong shock. Second, for the seed photons for IC emission, besides the cosmic microwave background (CMB), \citet{mezger1986maps,atoyan2000gamma} suggests an far-infrared (FIR) field with temperature $T_{\rm FIR}=97$ K, and energy density $U_{\rm FIR}=2\ \rm eV\ cm^{-3}$. Finally, as the magnetic field amplification is not well understood, we set the downstream magnetic fields for Syn emission as free parameters. However, as the magnetic field may be proportional to the shock velocity, we assume that the ratio between the downstream magnetic fields in the shocks of the jet and shell is \(B_{\rm d,jet}/B_{\rm d,shell}=v_{\rm jet}/v_{\rm shell}=6\) \citep{zhan2022asymmetrical}. 

We employ Naima \citep{naima}, a widely used Python package, for modeling non-thermal emissions, including Syn, IC and Bre. In addition, we apply aafragpy \citep{koldobskiy2021energy} in the calculation of PP collisions.

It is worth noting that in this physical model the free parameters are $B_d$, $A_e$, $A_p$, and $\alpha$, while the fixed parameters are $n_H$, $v_{\rm shell}$, $v_{\rm jet}$, $T_{\rm FIR}$, $U_{\rm FIR}$, and the parameters relevant to particle acceleration and escape timescales, i.e., $\eta_g$, $m$ and $\kappa$ (appendix ~\ref{app:timescale}).

\begin{table*}[ht]
\caption{{Spectral fitting results.}}
\centering
\begin{center}
\begin{tabular}{ccccccccccc}
\hline
\hline
 Model & Zone  &  ${B_d}$ &  $A_e$ & $A_p$ & ${\alpha}$ & ${E_{\rm cut,e}}$ & ${E_{\rm cut,p}}$  & ${E_b}$  & ${W_e}$ & ${W_p}$\\ 
   &   &  $\mathrm{(\mu G)}$ &  $\mathrm{(TeV^{-1})}$ & $\mathrm{(TeV^{-1})}$ &  & $\mathrm{(TeV)}$ & $\mathrm{(TeV)}$  & $\mathrm{(TeV)}$  & $\mathrm{(erg)}$ & $\mathrm{(erg)}$\\
 \hline 
M1  & shell & 121 & $5.37\times10^{46}$ & $3.24\times10^{48}$ & 2.43 & 21.58 & 45.88 & 2.52   & $7.60\times10^{49}$ &$1.73\times10^{50}$ \\ \hline
\multirow{2}{*}{M2}   & shell & 120 & $5.37\times10^{46}$ & $3.24\times10^{48}$ & 2.42 & 21.66 & 45.51 & 2.56 & $6.77\times10^{49}$ & $1.65\times10^{50}$ \\ 
  & jet & 720 & $3.98\times10^{44}$ &$2.34\times10^{46}$  & 2.42 & 53.24 & 3931 & 0.07 & $5.00\times10^{47}$ & $1.22\times10^{48}$ \\ \hline
\multirow{2}{*}{M3}   & shell & 119  &  $5.37\times10^{46}$ & $3.16\times10^{48}$ & 2.44 & 21.75 & 45.12 & 2.60 & $8.53\times10^{49}$ & $1.75\times10^{50}$ \\ 
  & jet & 714 & $2.00\times10^{44}$ & $2.51\times10^{46}$ & 2.00 & 53.46 & 3898 & 0.07 & $3.91\times10^{45}$ & $5.40\times10^{47}$ \\ \hline
\end{tabular}
\end{center}
\label{tab:results}
\end{table*}

\begin{figure}
\centering
\includegraphics[width=.47\textwidth]{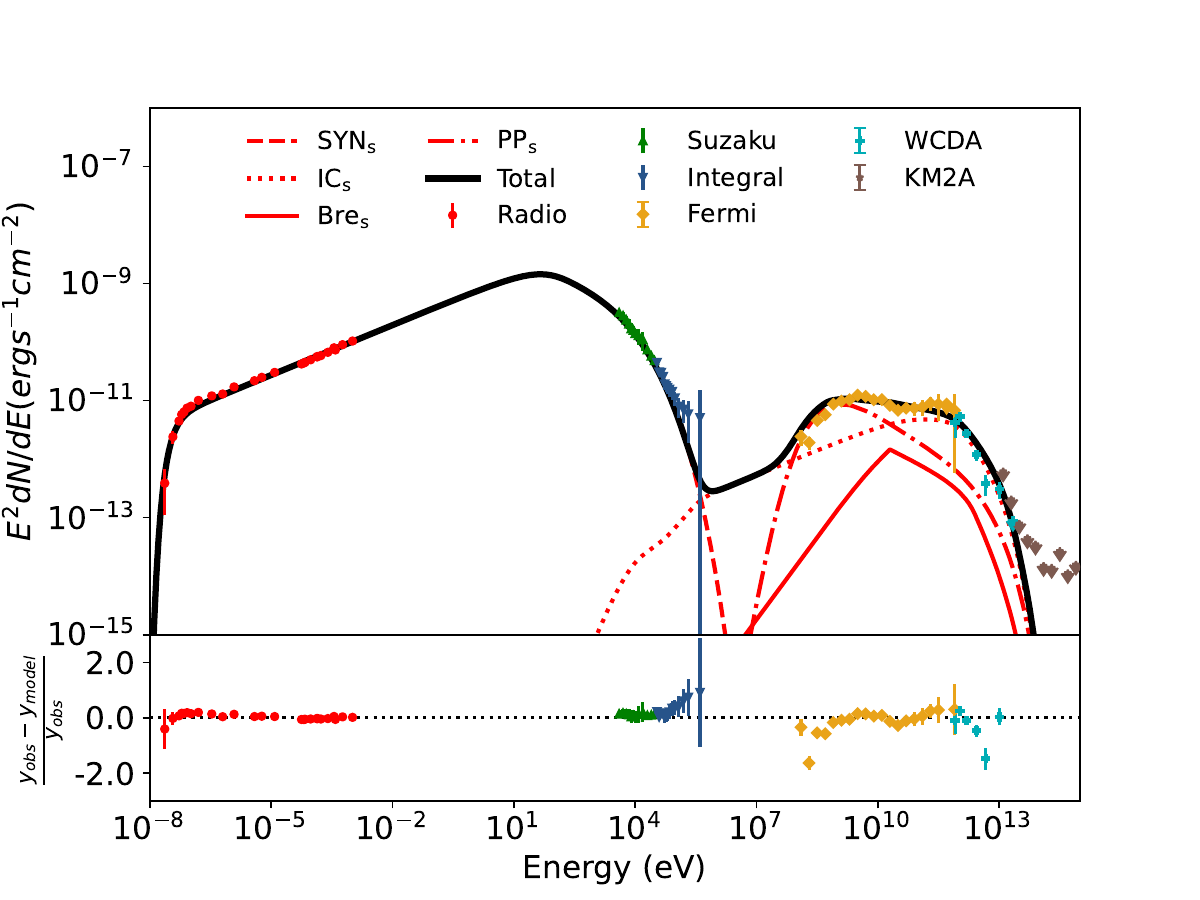}
\vfill
\includegraphics[width=.23\textwidth]{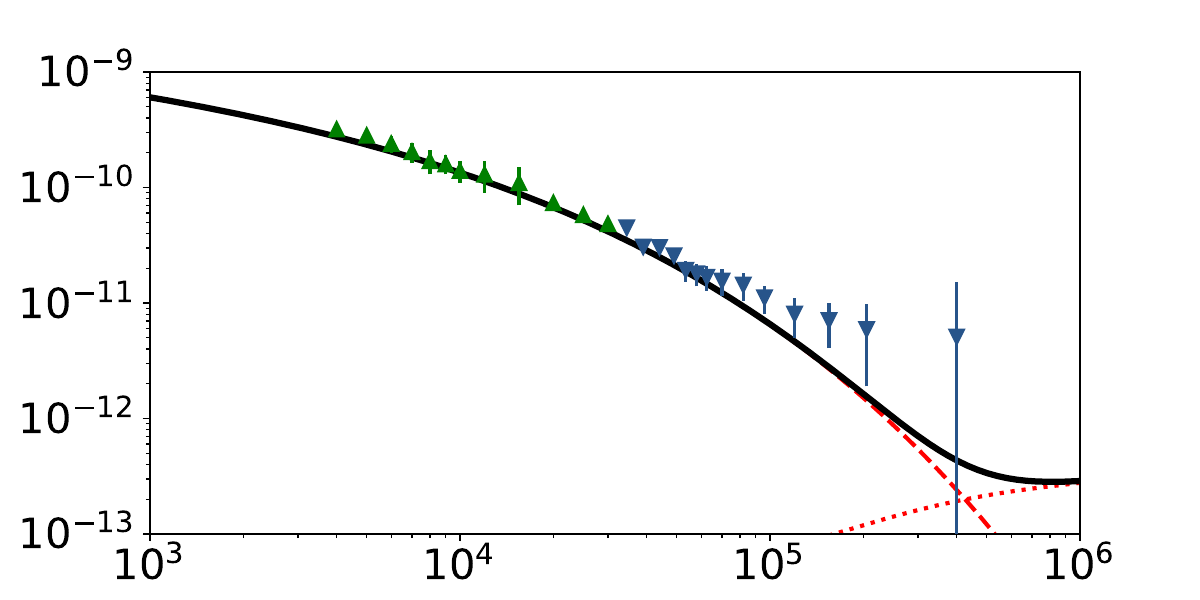}
\includegraphics[width=.23\textwidth]{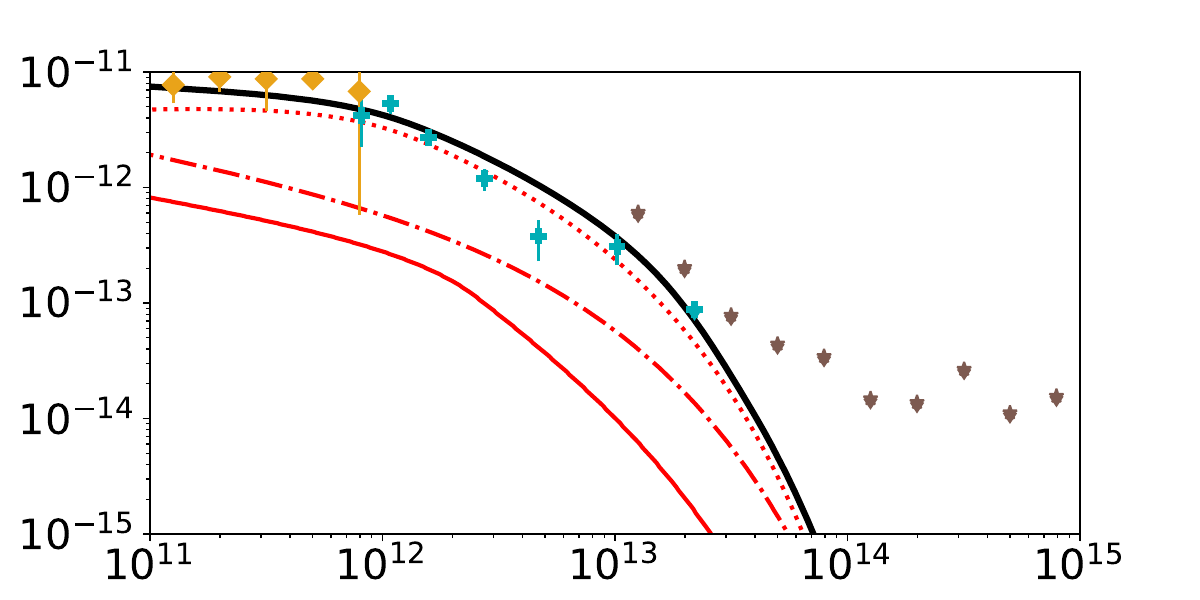}
\caption{Spectral fit with model M1. Upper panel: broadband spectrum and fitting results and relative error. Lower left panel: hard X-ray zoom-in. Lower right panel: gamma-ray zoom-in. }
\label{fig:M1}
\end{figure}

\begin{figure}
\centering
\includegraphics[width=.47\textwidth]{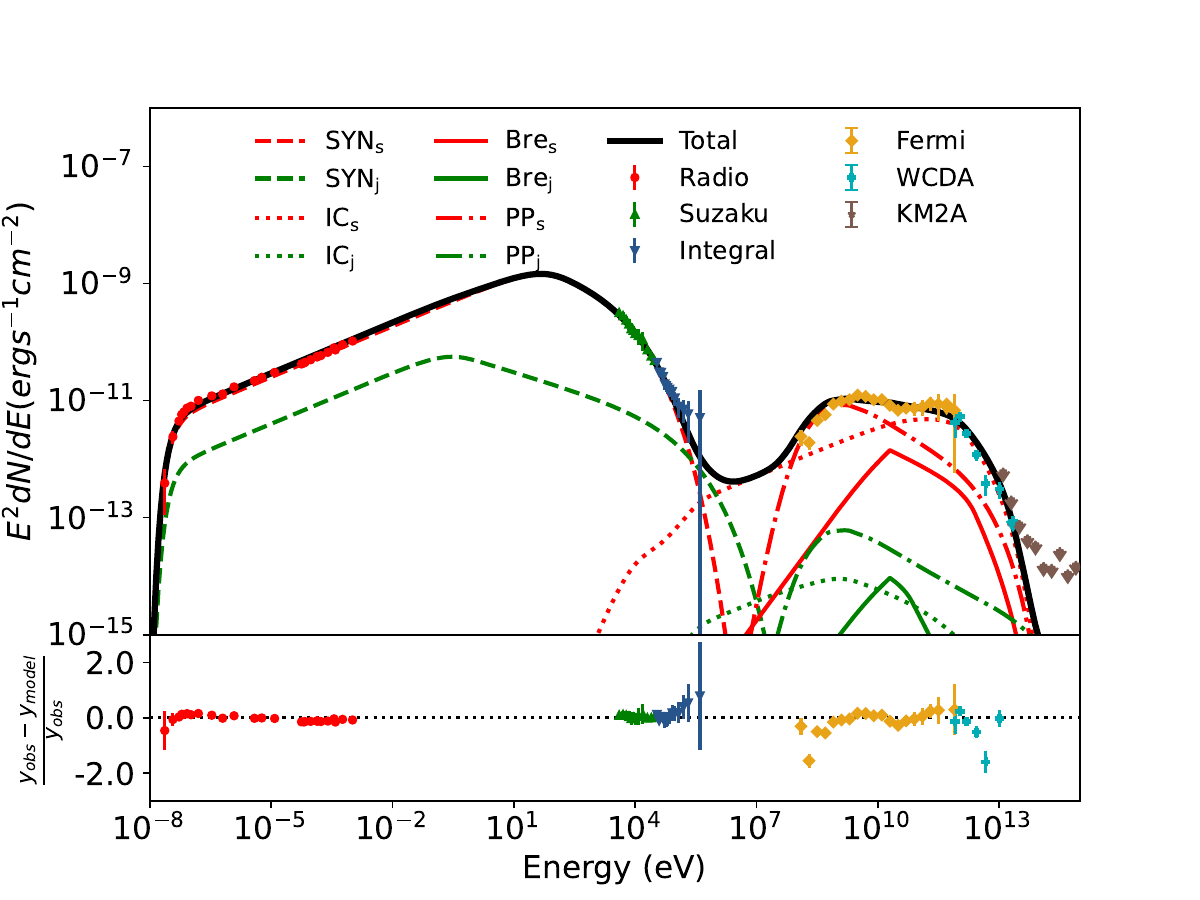}
\vfill
\includegraphics[width=.23\textwidth]{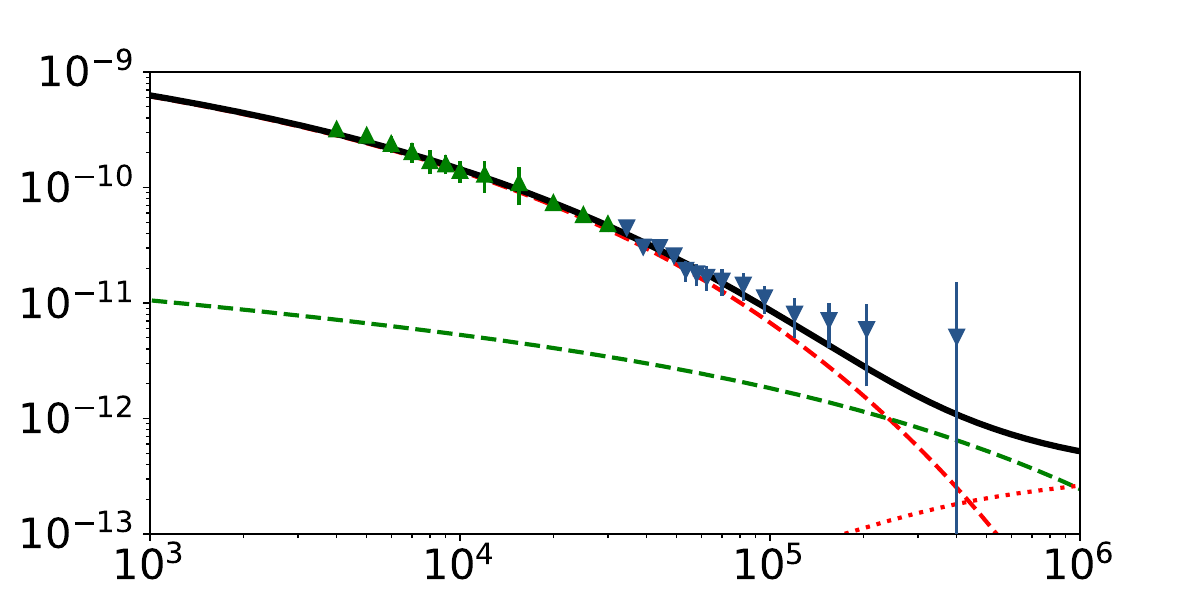}
\includegraphics[width=.23\textwidth]{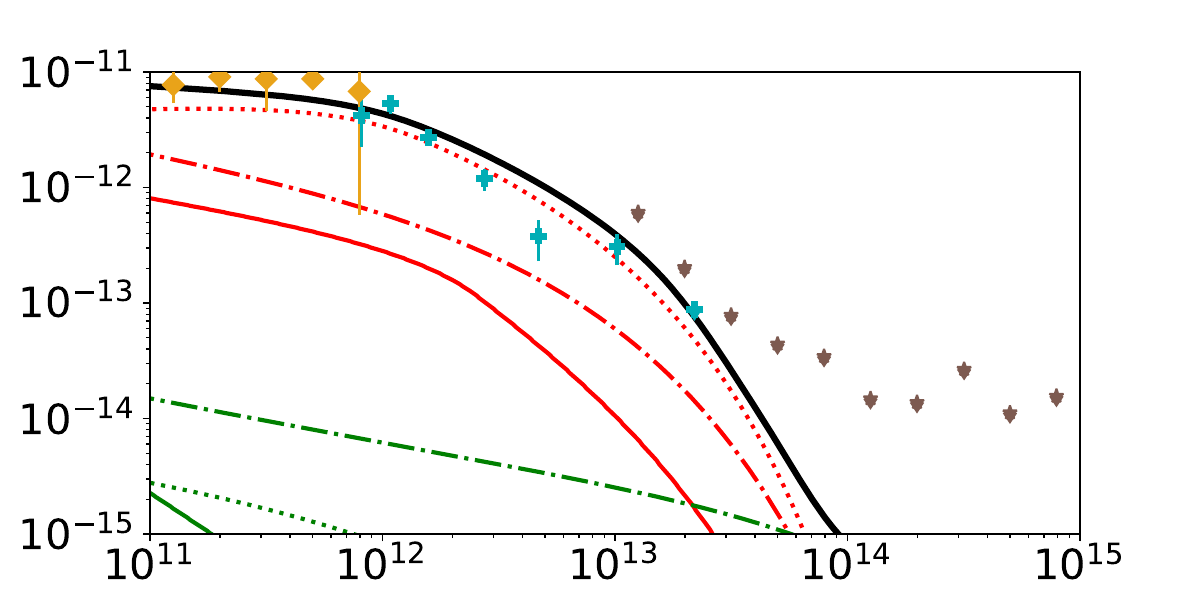}
\caption{Same as Figure~\ref{fig:M1}, but for model M2. Red and green lines present contribution from shell and jet zones, respectively. }  
\label{fig:M2}
\end{figure}

\begin{figure}
\centering
\includegraphics[width=.47\textwidth]{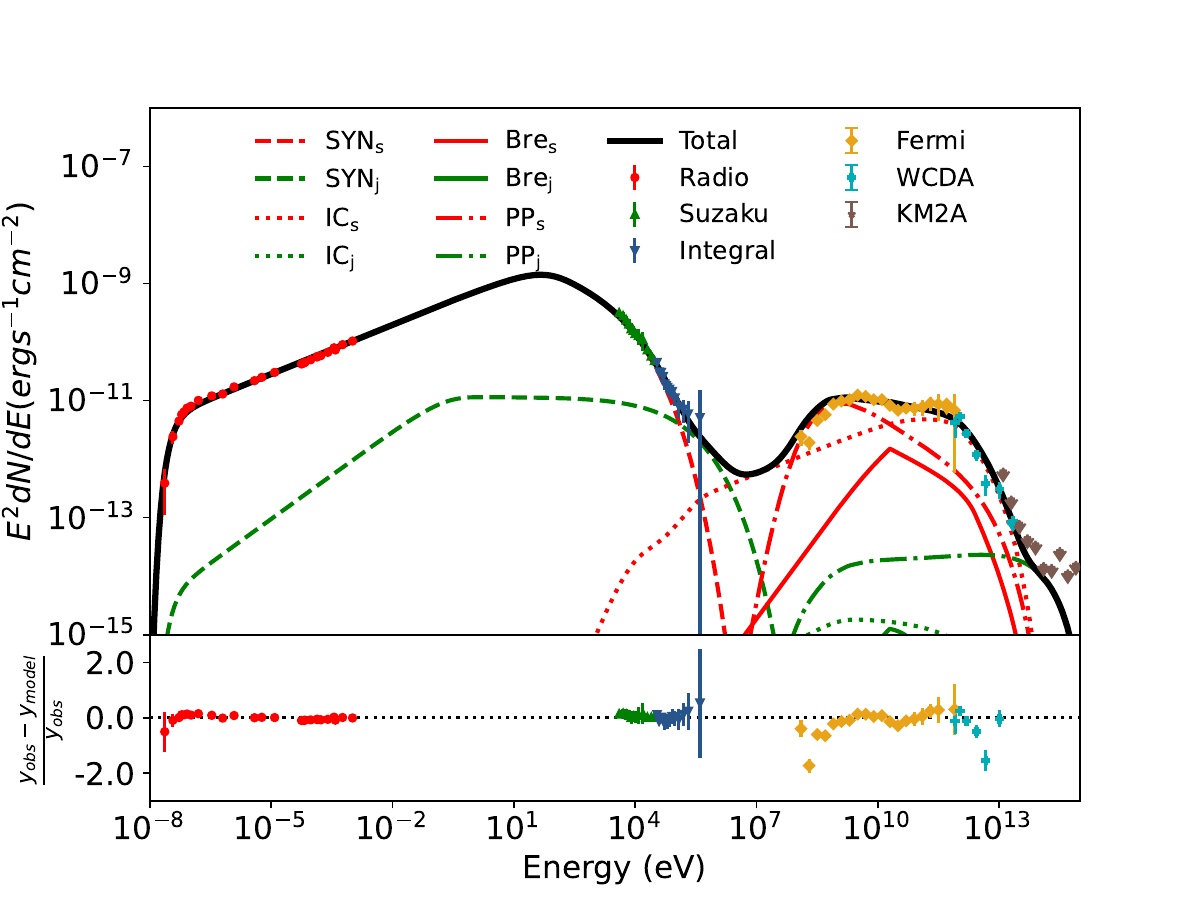}
\vfill
\includegraphics[width=.23\textwidth]{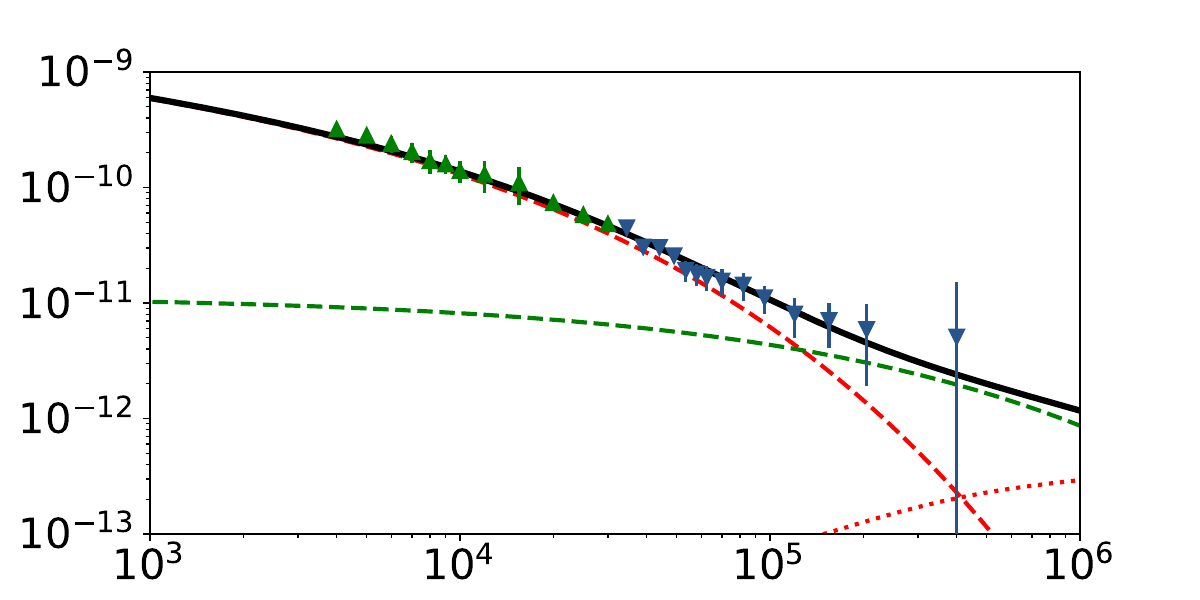}
\includegraphics[width=.23\textwidth]{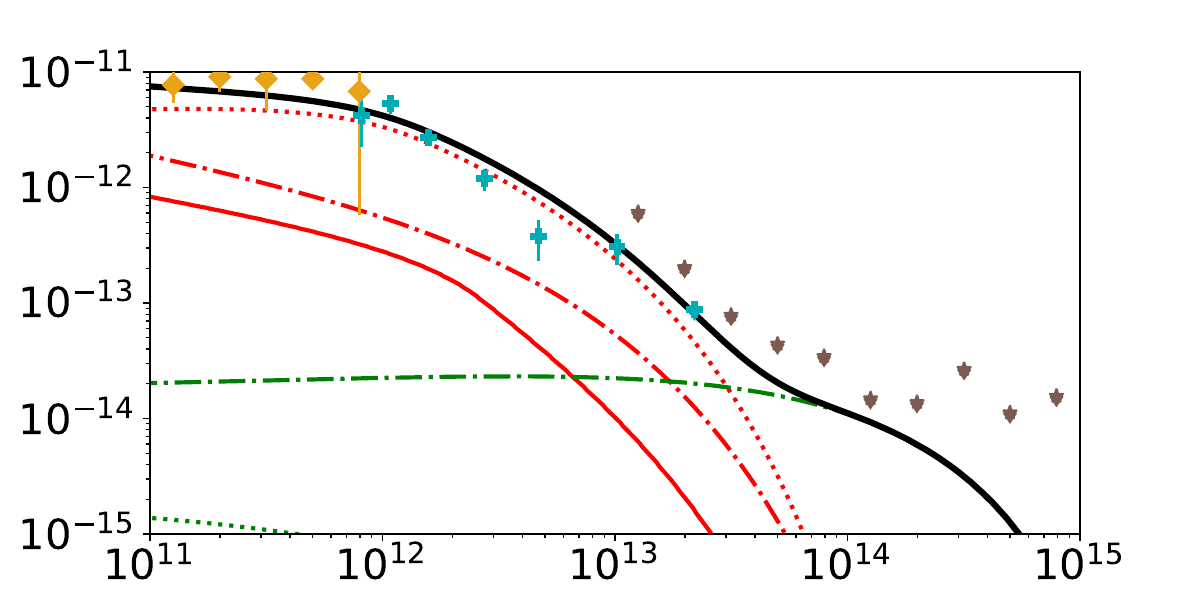}
\caption{Same as Figure~\ref{fig:M1}, but for model M3. Red and green lines present contribution from shell and jet zones, respectively.}
\label{fig:M3}
\end{figure}


\section{Fitting results} 
\label{sec:res}

Following \citet{zhan2022asymmetrical}, we consider three types of models in the broadband spectrum fitting: 
\begin{itemize}
    \item M1: a one-zone model without the jet structure; 
    \item M2: a shell-plus-jet two-zone model. This model assumes a similar acceleration for the shell and jet zones where they share the same particle indices $\alpha$ and electron-to-proton energy ratios  $W_e/W_p$. 
    \item M3: a shell-plus-jet two-zone model, where $\alpha$ and $W_e/W_p$ differ between two zones. 
\end{itemize}
The fitting results are shown in Table ~\ref{tab:results} and Figures ~\ref{fig:M1}, ~\ref{fig:M2}, and ~\ref{fig:M3}, where the fitting errors are calculated as relative error $[(y_{\rm obs}-y_{\rm model})/y_{\rm obs}]\times 100\%$. Note that in Table ~\ref{tab:results}, only $B_d$, $A_e$, $A_p$, and $\alpha$ are free parameters.

As shown in Figure ~\ref{fig:M1}, the one-zone model M1 well reproduces the observed emissions in radio, soft X-ray and gamma-ray bands. The radio and soft X-ray data constrain the spectral index to be $\alpha_{shell}\simeq2.42\pm0.02$. Moreover, in this model, the magnetic field can be well constrained too for that increasing the magnetic field will apply a complicated but important effect on the spectrum. Increasing $B_d$, on the one hand, will directly increase the Syn radiation and the acceleration efficiency. On the other hand, a stronger Syn cooling will lead to a lower break energy and cutoff energy, which will reduce IC emission and weaken the increase of Syn radiation in X-ray range. Therefore, a large field $B_d\sim\ 120\rm ~\mu G$ is constrained so that the model fits hard X-ray data as well as possible without exceeding the gamma-ray upper limit. 

However, model M1 faces significant challenges to account for the hard X-ray spectral tail above $\mathrm{50~ keV}$, as the DSA process predicts an exponential cutoff at a few keV naturally \citep{wang2016hard,zhan2022asymmetrical}. To fill these gaps, it is essential to involve additional physical processes capable of accelerating particles to higher energies and producing harder spectrum than the shell region. 

Next, an extra jet structure is included in two-zone models M2 and M3. In M2, the same index is adopted for jet and shell zones. As shown in Figure ~\ref{fig:M2}, the jet structure contributes an additional component in the hard X-ray band, providing a better fit compared to Figure ~\ref{fig:M1} where hard X-ray flux is underestimated. At the same time, the jet also leads to a weak but non-negligible contribution in the radio band, making M2 particularly sensitive to the magnetic field strength and spectral index.

The spectral index in M2 is slightly harder than that in M1 because jet zone contributes to radio and X-ray at the same time, requiring a harder shell component that would contributed less to radio band. The ratio $W_e/W_p=0.41$ in M2 is slightly higher than $W_e/W_p=0.439$ in M1. As Figure ~\ref{fig:M2} shows, the X-ray flux is obviously more consistent with the error bar. 

In M3, we relax the restriction in M2 by allowing different indices between jet and shell zones, as the different physical conditions may produce different particle spectra. We adopt the assumption of $\alpha=2$ for the jet zone, as this may be a reasonable result of the high velocity of the jet and hence a stronger shock. As shown in Figure ~\ref{fig:M3}, this hard acceleration results in a significant jet contribution in the hard X-ray above 100 keV, while still maintaining consistency with the uncertainties in 10-100 keV range. Additionally, the contribution of the jet's Syn radiation to radio band remains negligible for its hard spectrum. Thus, in M3, a softer shell index of $\alpha=2.44$ is applied.

Furthermore, Figure ~\ref{fig:M3} shows that PP collisions in jet zone would contribute to a sub-PeV gamma-ray tail of $\sim 1\times10^{-14}\rm erg\,cm^{-2}s^{-1}$ at 100 TeV in SED, satisfying the LHAASO-KM2A upper limits.

\section{Discussion}
\label{sec:dis}
\subsection{Model parameter uncertainties}
We have presented the fitting results of these three models for the broadband spectrum. The one-zone model provides a reasonable fit for most of the data, except for the hard X-rays above 50 keV, implying the necessity of an additional component with hard spectral tail. Adding a fast jet improves the overall fit. The observed spectrum from radio to X-ray bands is well reproduced by the two-zone model and generally lies within the uncertainties. The low-frequency radio emission below 100 MHz is affected by external absorption by the ISM so that the fitting in this part is less constrained.

\begin{figure*}[htbp]
\centering
\includegraphics[width=0.47\linewidth]{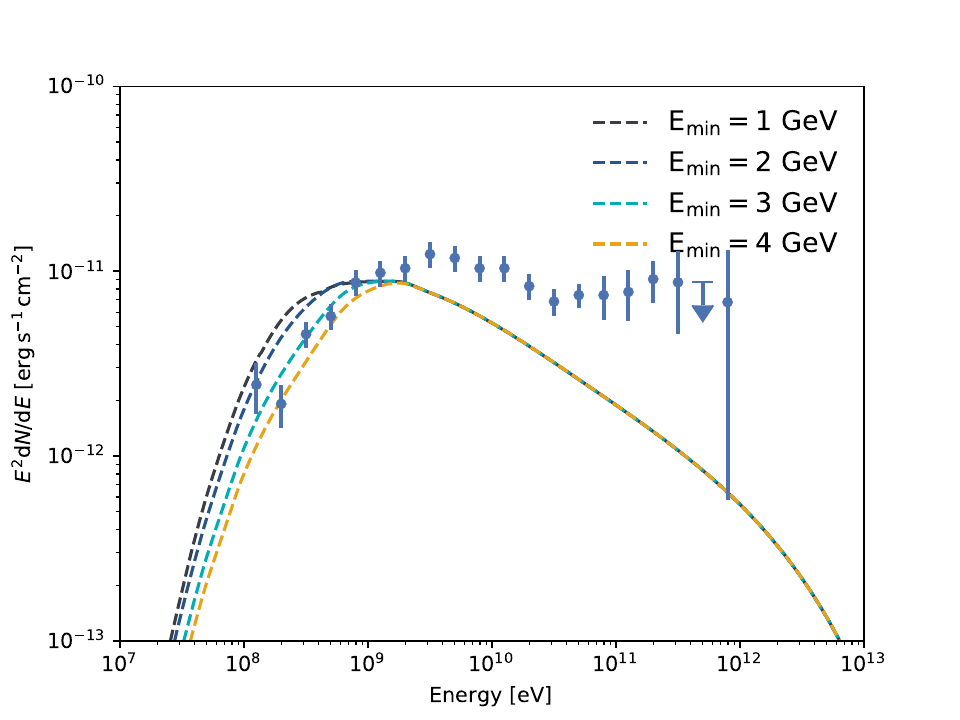}
\hfill
\includegraphics[width=0.47\linewidth]{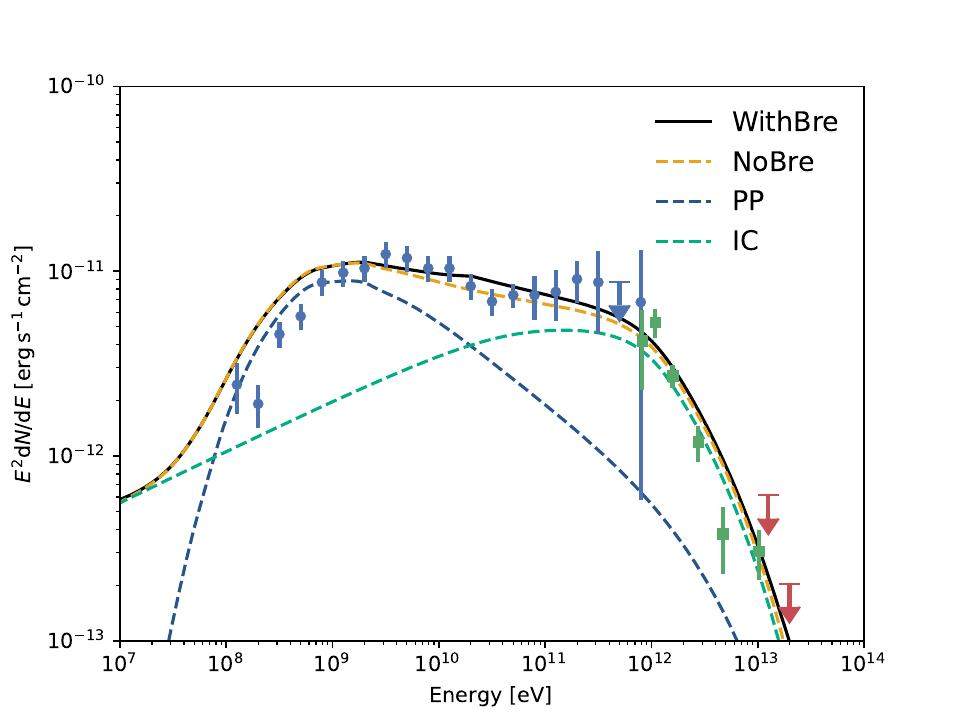}
\vspace{1em}
\includegraphics[width=0.47\linewidth]{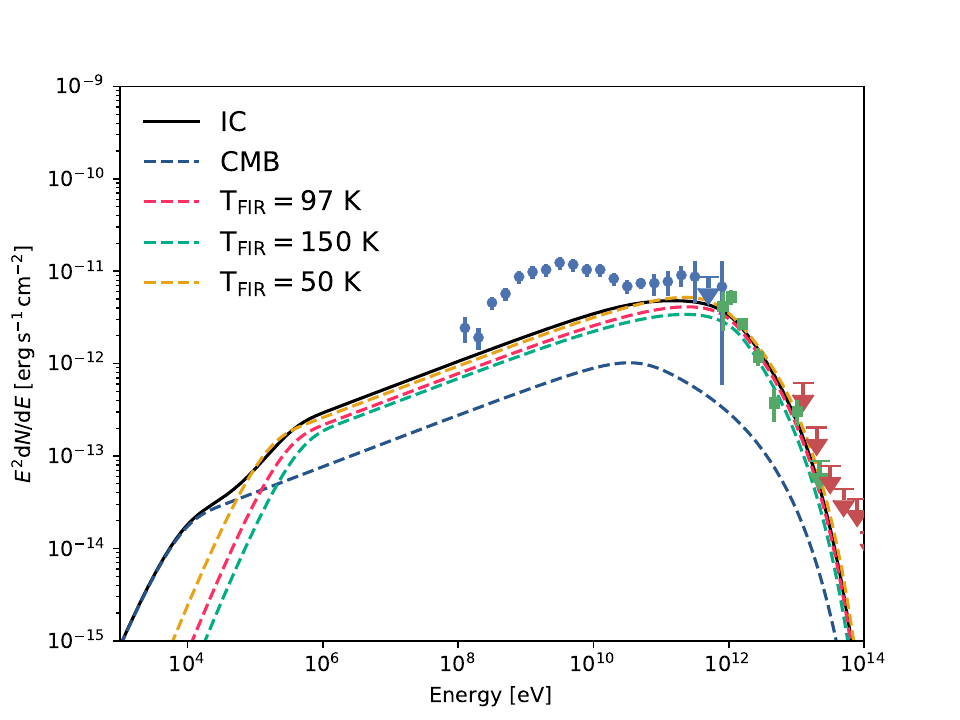}
\hfill
\includegraphics[width=0.47\linewidth]{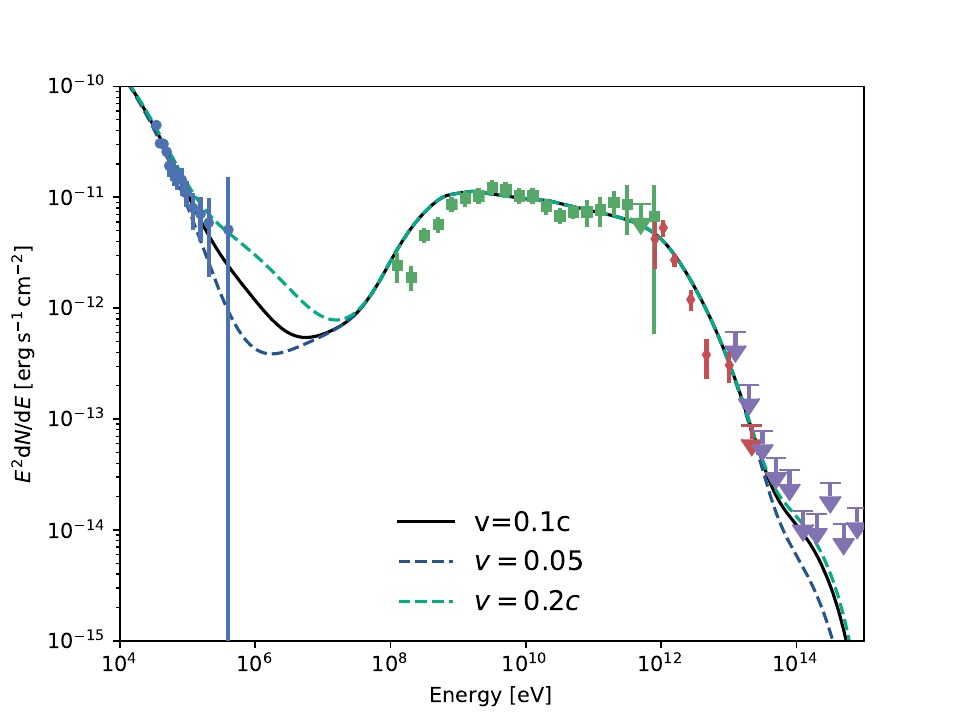}
\caption{GeV-TeV gamma-ray spectrum. Upper left panel: PP emissions with different lower energy limit assumotions. Upper right panel: SED with and without Bre emission. Lower left panel: IC emission with different FIR temperature assumptions. Lower right panel: SED for different jet velocity assumptions of 0.05c, 0.1c and 0.2c. }
\label{fig:gamma}
\end{figure*}

The gamma-ray emission from $\sim$100 MeV to 30 GeV is attributed to PP collisions. PP process can well account for the data, except for the data point at 200 MeV. Different lower limits of protons involved in the PP emission are shown in Figure ~\ref{fig:gamma}. As shown in this figure, a lower limit of 4 GeV fits the data better, but there is no theoretical basis to support the physical origin of this lower bound.

The spectral "dip" around 30 GeV is a strong evidence for the separate origins of spectral peaks around 3 GeV and 300 GeV, although our model expects a flatter spectrum between these two peaks. The peak of Bre emission, assuming background ion density $n_i=n_{\rm H}$, lies right around the dip, which erases somewhat the dip feature in the predicted spectrum. As there may be uncertainties in calculating Bre emission, e.g., $n_i$, the Bre emission in Cas A could be weaker than that predicted in M3. An example that sets Bre to be zero is shown in Figure ~\ref{fig:gamma}, where the double-peaked feature becomes more obvious. 

Above 1 TeV, our model suggests a spectrum dominated by IC emission. The calculated IC emission under different FIR field assumptions is shown in Figure ~\ref{fig:gamma}. In comparison, the observed spectrum declines more steeply than calculated in 1–5 TeV range, but seems to hint at a spectral hardening near 10 TeV. This may be partially explained by a broken power law with a much faster spectral index jump. Clearly explaining this part would be difficult. Nevertheless, our model remains within a $5\sigma$ confidence level and attributes the sharp 1-5 TeV decline to limitations in sensitivity and statistics. Beyond 10 TeV, our model is within the allowable upper limit. 

We note some other parameter uncertainties here. The jet velocity of $30000\rm km\,s^{-1}$ is only an assumption because only transverse velocity $14000\rm km\,s^{-1}$ is measured. Different assumptions for velocities of $14000\rm km\,s^{-1}$ and $60000 \rm km\,s^{-1}$ are modeled and shown in Figure ~\ref{fig:gamma}. Indices between $\alpha=2$ and 2.1 yield comparable good fits to the hard X-ray and are consistent with the gamma-ray upper limits at the same time. In fact a harder spectrum gives a better fitting result to the hard X-ray emission naturally. In addition, we have fixed those parameters relevant to particle diffusion and escaping, i.e., $\eta_g$ and $\kappa$ \citep{zhan2022asymmetrical}, within the uncertainties of DSA theory. Their values mainly affect the maximum energies of the accelerated particles. For example, here we take $\eta_g=1.3$, instead of $\eta_g=2.2$ in \cite{zhan2022asymmetrical}, so as to adjust the maximum particle energy to match the new TeV data from LHAASO.

\subsection{Comparison with previous studies}
Here, we compare our results with previous studies on gamma-ray spectrum of Cas A. As \citet{ahnen2017cut,abeysekara2020evidence} puts it, MAGIC and VERITAS measured the gamma-ray spectrum from 0.1 GeV to 10 TeV, suggesting a spectral cutoff at $\mathrm{2-3TeV}$ with an index of $\alpha=2.4\pm0.3$. The low flux measured by MAGIC and VERITAS is inconsistent with the 0.3 TeV peak detected by Fermi-LAT. This results in the spectrum being considered as single-peaked in previous works. However, LHAASO observations, along with 16 years of Fermi-LAT observations, present a larger flux than that observed by MAGIC and VERITAS \citep{cao2025broadband}, as shown in Figure ~\ref{fig:spec}, clearly indicating a double-peaked structure.  

Previous studies generally considered a spectrum with a single peak around 10 GeV, assuming a unified origin for 100 MeV-30 GeV and 40 GeV-10 TeV emissions. In the VERITAS collaboration's paper, they applied a purely hadronic model with $B_{\min}=450\ \mu G$ to suppress IC contribution and a lepto-hadronic model with $B_{\min}=150\ \mu G$ for IC emission to account for energy above 1 TeV \citep{abeysekara2020evidence}. This result agrees with our model, because a double-peaked model requires a larger break energy and thus implies a smaller $B_{d}\sim120\mu G$ for the SED of leptonic emission to account for the peak at $\sim$ 0.3 TeV. 

Both the MAGIC \citep{ahnen2017cut} and VERITAS's \citep{abeysekara2020evidence} studies conclude that Cas A can only accelerate CRs to TeV scale, based on the observed TeV cutoff. This is in fact consistent with our model in which this TeV cutoff takes place in the shell zone, while the jet can still accelerate protons to PeV, producing sub-PeV gamma-rays via PP collisions.

As discussed earlier, a one-zone scenario is insufficient to account for the broadband spectrum of Cas A. Based on the radio and X-ray spectra and imaging, a two-zone model separating the low and high density regions, corresponding to a flat and a steep spectral component in radio band, was proposed by \cite{atoyan2000energy}. \cite{zhang2019supernova} also propose a two-zone model similar to that of \cite{atoyan2000energy}, but regard the two zones as the forward and reverse shock regions. They treat the maximum energy of particles as free parameters in spectral fitting, whereas we adopt a more physical approach to determine it. Notably, in their spectral fitting they require a Syn spectral cutoff energy in the reverse shock larger than that in the forward shock, in contrast with the prediction by the DSA that the Syn cutoff is proportional to shock velocity squared, $\propto v_s^2$, since the reverse shock velocity is usually smaller than the forward shock. However, this is supported by that X-ray knots are observed in the inner region of Cas A with an inward velocity relative to the ejecta as high as 8700 $\rm km\,s^{-1}$ \citep{sato2018ApJ...853...46S}. In this scenario, the hard X-ray emission is suggested to be produced by Bre from a group of non-thermal and NR electrons \citep[see also][]{cao2025broadband}. 

Compared with previous results from \citep{zhan2022asymmetrical}, the new Fermi-LAT and LHAASO data lead to a different and improved fit. In this paper, we take $\eta_g=1.3$, instead of $\eta_g=2.2$. A relativistic condition of PP collisions and a harder spectrum assumption are used to shift the PP peak to a higher energy range. Moreover, new LHAASO-KM2A data indicate a lower upper bound, thus a lower jet power and particles' population. 

\subsection{Cas A-like SNRs as PeVatrons?}
The results of our model suggest that the jet of Cas A can reasonably accelerate CR protons up to a few PeV for a high jet velocity of $v_{\rm jet}\sim0.1c$, but the total proton energy in the jet is constrained to be around $W_p=5\times10^{47}$ erg for accelerated protons with energies from $\sim2$ GeV to $\sim2$ PeV. Can SNRs like Cas A produce enough amount of CRs to account for the observed PeV CRs?

If the event rate of Cas A-like supernovae in our Galaxy is $R_{SN}\simeq10^{-2}\rm yr^{-1}$ \citep[e.g.,][]{2013MNRAS.435.1174S}, this corresponds to a CR energy production rate in the range of $E=2\rm GeV-2PeV$ of $Q=W_pR_{SN}=1.6\times10^{38}\rm erg\,s^{-1}$. For spectral index $\alpha=2$ (Model M3 in Table \ref{tab:results}), the differential production rate is $E^2Q(E)=1.2\times10^{37}\rm erg\,s^{-1}$, and $Q(E=1{\rm PeV})=7.2\times10^{27}\rm s^{-1}GeV^{-1}$. The CR intensity is $J(E)=(c/4\pi)Q(E)t_{conf}(E)/V$, with $t_{conf}$ being the CR confinement time in the Galaxy, and $V$ being the volume of the Galaxy. Taking $V=\pi r_G^2h_G$, with $r_G=10$ kpc and $h_G=1$ kpc as the radius and CR height of the Galactic plane, and using the observed CR intensity at $E=1$ PeV, $J(E)=9\times10^{-17}\rm cm^{-2}s^{-1}sr^{-1}GeV^{-1}$ \citep{youzhiyong}, the confinement time is derived to be $t_{conf}(E=1{\rm PeV})=1.5(h_G/1\rm\, kpc)$ Myr. This seems too large, almost comparable to the measured confinement time of GeV CRs, as PeV CRs propagating much faster than GeV CRs should have much smaller confinement time.

However, for high-enough-energy CRs, as they propagate out from the Galactic plane quickly, CRs may only occupy the local region around the sources, say, the Cas A-like SNRs, so that the CR spatial distribution in the Galaxy could be clumpy. In this case the CR intensity in a CR clump is $J(E)=(c/4\pi)EN_p(E)/V_c$ on average, where $V_c$ is the clump volume. For $V_c\sim(4/3)\pi r_c^3$, and using the observed CR proton flux at $E=1$ PeV, we derive the clump radius of PeV CRs, $r_c\simeq0.2$ kpc, which is comparable to half of the thickness of the Galaxy plane. Thus, it is likely that the spatial distribution of PeV CRs is clumpy in the Galaxy and the Earth is located in a CR clump with relatively large PeV CR density. 

The derivation of the clump volume is independent of $R_{SN}$ and $t_{conf}$. However, to be clumpy, the clump volumes should not overlap with each other. As SNRs occur in the Galaxy disk, this requires that the clump number satisfies $N_c=R_{SN}t_{conf}<(r_G/r_c)^2$, which reduces to a constraint of the confinement time, $t_{conf}<0.3$ Myr for $E=1$ PeV. Thus, we encourage further long-term monitoring of Cas A using LHAASO and CTAO to better constrain the sub-PeV spectrum. The existence of sub-PeV emission would support the idea that Cas A is able to accelerate protons to PeV energy in the fast jet. 

\section{Conclusion}
\label{sec:con}

In this work, we use an asymmetric explosion model to interpret the broadband spectrum of Cas A, which introduces a fast jet structure with a velocity of $\mathrm{30000\ km\ s^{-1}}$, in addition to the traditional shell structure \citep{wang2016hard,zhan2022asymmetrical}. This model successfully explains the absence of spectral cutoff in hard X-rays above 50 keV and still predicts the possibly existing sub-PeV gamma-ray tail. 

Our spectrum fitting results show that Syn emission from the shell zone can account for the data from radio to soft X-ray band, while Syn emission from the jet contributes to the hard X-ray spectrum. The magnetic field strength in the shell is constrained to $\sim 120\mu G$ for its dominant role in deciding break energy and maximum energy through characteristic timescales. Meanwhile, the shell component can account for the gamma-ray spectrum from GeV to TeV with a PP component dominating the GeV band and an IC component dominating the TeV band.

Unlike previous studies that disfavors Cas A as a PeVatron candidate due to the observed TeV cutoff \citep{guberman2017deep,cao2024does,cao2025broadband}, our results suggest that the fast-speed jet may be able to accelerate protons to PeV energy due to its extreme acceleration efficiency. The Cas A-like SNRs could be the Galactic PeVatrons if the PeV CR distribution is clumpy in the Galaxy with the size of the clump comparable to the thickness of the Galaxy. Further observations by LHAASO and CTAO will be crucial to test this model. Confirmation of sub-PeV emission would provide direct evidence for the presence of PeV protons from the jet. 

We note that the most significant contrast of the single-zone model with observations lies in the hard X-ray band. The X-ray cutoff detected by Suzaku at a few keV agrees well with the DSA expectation, and its spectrum from 20 to 40 keV is consistent with the INTEGRAL measurements. However, the INTEGRAL spectrum from 80 to 220 keV decreases much more slowly than DSA expectation, suggesting the presence of an additional hard spectral component at high energies. Many possible origins have been discussed and excluded but a high-speed jet model is favored \citep{wang2016hard}. However, as the DSA mechanism is not fully understood, it is not ruled out that the electron distribution declines gradually rather than exponentially (Eq. \ref{eq:ecbpl}), such that the SNR shell itself may produce the hard spectral component. Nevertheless, the existence of jets in Cas A has been evidenced by imaging and proper-motion measurements \citep{hwang2004million,fesen2006expansion}, and supported by observed $^{44}$Ti yield \citep{wang2016hard}. The jet contribution is not in contrast with the observed broadband spectrum, and may release the tension with hard X-ray measurements. 



\section*{Data Availability}
Data that were used in this paper are taken from previously published papers. Programs used like naima and aafragpy are available in github.

\begin{acknowledgments}
This work is supported by the National Key R\&D program of China (under the grant 2024YFA1611402, 2021YFA0718503 and 2023YFA1607901) and Natural Science Foundation of China (No. 12575116, 12133007).
\end{acknowledgments}

\appendix
\section{Time scales}\label{app:timescale}

The relativistic particle diffusion coefficient is given as $D=(1/3)lc$. The Bohm limit $D_B$ is the case that the mean free path is the Larmor radius, $l=E/eB$. General diffusion cofficient is usually assumed to be $D=\eta_gD_B$, with $\eta_g\sim$a few. The acceleration time scale of particles by strong shock can be given as,
\begin{equation}
    t_{\rm acc}\approx\frac{20D}{v_s^2}=\eta_g\frac{20D_B}{v_s^2}=\eta_g\frac{20Ec}{3eB_uv_s^2}
\end{equation}
with $v_s$ the shock velocity, and $B_u$ the upstream field. We should assume that the ratio of downstream and upstream magnetic field is $B_d/B_u\simeq m\sim3$ (see discussion in Ref. \citep{zhan2022asymmetrical}) due to magnetic field amplification. For typical values, one has
\begin{equation}
    t_{\rm acc}=2.1\times 10^3\eta_g\frac{E}{1\rm TeV}\left(\frac{B_u}{1\mu \rm G}\right)^{-1}\left(\frac{v_s}{10^8\rm cm\,s^{-1}}\right)^{-2}\rm yr.
\end{equation}

If the free escape boundary upstream is set to be $\kappa R$ ahead of the shock radius $R$, the escape time scale is the time that particles propagate such a distance, 
\begin{equation}
    \tau_{\rm esc}=\frac{(\kappa R)^2}{D}=\frac{(\kappa R)^2}{\eta_gD_B}=\kappa^2\frac{3eB_uR^2}{\eta_gEc},
\end{equation}
with unknown parameter $\kappa\sim0.1$. We have
\begin{equation}
\tau_{\rm esc}= 9\times 10^3 \kappa^2\eta_{\rm g}^{-1}\left(\frac{E}{1\rm TeV}\right)^{-1} \frac{B_u}{1\rm \mu G} \left(\frac{R}{1\rm pc}\right)^2{\rm yr}.
\end{equation}
In Cas A, we take $R=2.5$pc for SNR shell and $R=6$pc for the jet, according to the X-ray image and taking into account the projection effect. And we will fix the uncertain parameters as $m=2.1$, $\eta_g=1.3$ and $\kappa=0.04$.

For energy loss of particles, we have the Syn, IC and Bre cooling timescale of electrons, and the PP cooling timescales as below \citep{zhan2022asymmetrical}:
\begin{equation}\label{eq:syn}
\tau_{\rm syn} = 1.3\times10^{7} \left(\frac{E}{1\rm TeV}\right)^{-1} \left(\frac{B_{d}}{1 \rm \mu G}\right)^{-2}\rm yr ,
\end{equation}
\begin{equation}
\tau_{\rm IC} = 3.1\times10^5 \left(\frac{E}{1\rm TeV} \right)^{-1}\left(\frac{u_{\rm ph}}{\rm 1\, eV \ cm^{-3}}\right)^{-1}\rm yr,
\end{equation}
\begin{equation}
\tau_{\rm bre}=4\times10^{7} \left(\frac{n_{\rm H}}{1\rm cm^{-3}}\right)^{-1}\rm yr,
\end{equation}
\begin{equation}
\tau_{\rm pp}=6\times10^{7}\left(\frac{n_{\rm H}}{1{\rm cm}^{-3}}\right)^{-1} {\rm yr}, 
\end{equation}
where $u_{\rm ph}$ is the radiation energy density, and $n_{\rm H}$ is the downstream particle number density.

\bibliography{bib}

@ARTICLE{2008ApJ...678..939Z,
       author = {{Zirakashvili}, V.~N. and {Ptuskin}, V.~S.},
        title = "{Diffusive Shock Acceleration with Magnetic Amplification by Nonresonant Streaming Instability in Supernova Remnants}",
      journal = {\apj},
         year = 2008,
        month = may,
       volume = {678},
       number = {2},
        pages = {939-949},
          doi = {10.1086/529580},
archivePrefix = {arXiv},
       eprint = {0801.4488},
 primaryClass = {astro-ph},
       adsurl = {https://ui.adsabs.harvard.edu/abs/2008ApJ...678..939Z}
}

@ARTICLE{2011ApJ...731...87E,
       author = {{Ellison}, Donald C. and {Bykov}, Andrei M.},
        title = "{Gamma-ray Emission of Accelerated Particles Escaping a Supernova Remnant in a Molecular Cloud}",
      journal = {\apj},
         year = 2011,
        month = apr,
       volume = {731},
       number = {2},
          eid = {87},
        pages = {87},
          doi = {10.1088/0004-637X/731/2/87},
archivePrefix = {arXiv},
       eprint = {1102.3885},
 primaryClass = {astro-ph.HE},
       adsurl = {https://ui.adsabs.harvard.edu/abs/2011ApJ...731...87E}
}

@ARTICLE{youzhiyong,
       author = {{The LHAASO Collaboration} and {Cao}, Zhen and {Aharonian}, F. and {Bai}, Y.~X. and {Bao}, Y.~W. and {Bastieri}, D. and {Bi}, X.~J. and {Bi}, Y.~J. and {Bian}, W. and others},
        title = "{First Identification and Precise Spectral Measurement of the Proton Component in the Cosmic-Ray `Knee'}",
      journal = {arXiv e-prints},
         year = 2025,
        month = may,
          eid = {arXiv:2505.14447},
        pages = {arXiv:2505.14447},
          doi = {10.48550/arXiv.2505.14447},
archivePrefix = {arXiv},
       eprint = {2505.14447},
 primaryClass = {astro-ph.HE},
       adsurl = {https://ui.adsabs.harvard.edu/abs/2025arXiv250514447T}
}

@ARTICLE{2012APh....39...52D,
       author = {{Drury}, Luke O. 'C.},
        title = "{Origin of cosmic rays}",
      journal = {Astroparticle Physics},
         year = 2012,
        month = dec,
       volume = {39},
        pages = {52-60},
          doi = {10.1016/j.astropartphys.2012.02.006},
archivePrefix = {arXiv},
       eprint = {1203.3681},
 primaryClass = {astro-ph.HE},
       adsurl = {https://ui.adsabs.harvard.edu/abs/2012APh....39...52D}
}

@ARTICLE{2013MNRAS.435.1174S,
       author = {{Schure}, K.~M. and {Bell}, A.~R.},
        title = "{Cosmic ray acceleration in young supernova remnants}",
      journal = {\mnras},
         year = 2013,
        month = oct,
       volume = {435},
       number = {2},
        pages = {1174-1185},
          doi = {10.1093/mnras/stt1371},
archivePrefix = {arXiv},
       eprint = {1307.6575},
 primaryClass = {astro-ph.HE},
       adsurl = {https://ui.adsabs.harvard.edu/abs/2013MNRAS.435.1174S}
}

@ARTICLE{2014ApJ...785..130Z,
       author = {{Zirakashvili}, V.~N. and {Aharonian}, F.~A. and {Yang}, R. and {O{\~n}a-Wilhelmi}, E. and {Tuffs}, R.~J.},
        title = "{Nonthermal Radiation of Young Supernova Remnants: The Case of CAS A}",
      journal = {\apj},
         year = 2014,
        month = apr,
       volume = {785},
       number = {2},
          eid = {130},
        pages = {130},
          doi = {10.1088/0004-637X/785/2/130},
archivePrefix = {arXiv},
       eprint = {1308.3742},
 primaryClass = {astro-ph.HE},
       adsurl = {https://ui.adsabs.harvard.edu/abs/2014ApJ...785..130Z}
}

@ARTICLE{sato2018ApJ...853...46S,
       author = {{Sato}, Toshiki and {Katsuda}, Satoru and {Morii}, Mikio and {Bamba}, Aya and {Hughes}, John P. and {Maeda}, Yoshitomo and {Ishida}, Manabu and {Fraschetti}, Federico},
        title = "{X-Ray Measurements of the Particle Acceleration Properties at Inward Shocks in Cassiopeia A}",
      journal = {\apj},
         year = 2018,
        month = jan,
       volume = {853},
       number = {1},
          eid = {46},
        pages = {46},
          doi = {10.3847/1538-4357/aaa021},
archivePrefix = {arXiv},
       eprint = {1710.06992},
 primaryClass = {astro-ph.HE},
       adsurl = {https://ui.adsabs.harvard.edu/abs/2018ApJ...853...46S}
}

@ARTICLE{2010ApJ...720...20A,
       author = {{Araya}, Miguel and {Cui}, Wei},
        title = "{Evidence for Cosmic Ray Acceleration in Cassiopeia A}",
      journal = {\apj},
         year = 2010,
        month = sep,
       volume = {720},
       number = {1},
        pages = {20-25},
          doi = {10.1088/0004-637X/720/1/20},
archivePrefix = {arXiv},
       eprint = {1006.5962},
 primaryClass = {astro-ph.HE},
       adsurl = {https://ui.adsabs.harvard.edu/abs/2010ApJ...720...20A}
}

@ARTICLE{1978ApJ...221L..29B,
       author = {{Blandford}, R.~D. and {Ostriker}, J.~P.},
        title = "{Particle acceleration by astrophysical shocks.}",
      journal = {\apjl},
         year = 1978,
        month = apr,
       volume = {221},
        pages = {L29-L32},
          doi = {10.1086/182658},
       adsurl = {https://ui.adsabs.harvard.edu/abs/1978ApJ...221L..29B}
}

@ARTICLE{2000ApJ...542..235K,
       author = {{Kirk}, J.~G. and {Guthmann}, A.~W. and {Gallant}, Y.~A. and {Achterberg}, A.},
        title = "{Particle Acceleration at Ultrarelativistic Shocks: An Eigenfunction Method}",
      journal = {\apj},
         year = 2000,
        month = oct,
       volume = {542},
       number = {1},
        pages = {235-242},
          doi = {10.1086/309533},
archivePrefix = {arXiv},
       eprint = {astro-ph/0005222},
 primaryClass = {astro-ph},
       adsurl = {https://ui.adsabs.harvard.edu/abs/2000ApJ...542..235K}
}

@ARTICLE{2005PhRvL..94k1102K,
       author = {{Keshet}, Uri and {Waxman}, Eli},
        title = "{Energy Spectrum of Particles Accelerated in Relativistic Collisionless Shocks}",
      journal = {\prl},
         year = 2005,
        month = mar,
       volume = {94},
       number = {11},
          eid = {111102},
        pages = {111102},
          doi = {10.1103/PhysRevLett.94.111102},
archivePrefix = {arXiv},
       eprint = {astro-ph/0408489},
 primaryClass = {astro-ph},
       adsurl = {https://ui.adsabs.harvard.edu/abs/2005PhRvL..94k1102K}
}

@ARTICLE{1987PhR...154....1B,
       author = {{Blandford}, Roger and {Eichler}, David},
        title = "{Particle acceleration at astrophysical shocks: A theory of cosmic ray origin}",
      journal = {\physrep},
         year = 1987,
        month = oct,
       volume = {154},
       number = {1},
        pages = {1-75},
          doi = {10.1016/0370-1573(87)90134-7},
       adsurl = {https://ui.adsabs.harvard.edu/abs/1987PhR...154....1B}
}

@ARTICLE{2013ApJ...778..107S,
       author = {{Sironi}, Lorenzo and {Giannios}, Dimitrios},
        title = "{A Late-time Flattening of Light Curves in Gamma-Ray Burst Afterglows}",
      journal = {\apj},
         year = 2013,
        month = dec,
       volume = {778},
       number = {2},
          eid = {107},
        pages = {107},
          doi = {10.1088/0004-637X/778/2/107},
archivePrefix = {arXiv},
       eprint = {1307.3250},
 primaryClass = {astro-ph.HE},
       adsurl = {https://ui.adsabs.harvard.edu/abs/2013ApJ...778..107S}
}

@ARTICLE{2018ApJ...859..123H,
       author = {{Huang}, Yan and {Li}, Zhuo},
        title = "{Persistent X-Ray Emission from ASASSN-15lh: Massive Ejecta and Pre-SLSN Dense Wind?}",
      journal = {\apj},
         year = 2018,
        month = jun,
       volume = {859},
       number = {2},
          eid = {123},
        pages = {123},
          doi = {10.3847/1538-4357/aabcca},
archivePrefix = {arXiv},
       eprint = {1801.07517},
 primaryClass = {astro-ph.HE},
       adsurl = {https://ui.adsabs.harvard.edu/abs/2018ApJ...859..123H}
}

@inproceedings{krymskii1977regular,
  title={A regular mechanism for the acceleration of charged particles on the front of a shock wave},
  author={Krymskii, GF},
  booktitle={Akademiia Nauk SSSR Doklady},
  volume={234},
  pages={1306--1308},
  year={1977}
}

@article{bell1978acceleration,
  title={The acceleration of cosmic rays in shock fronts--I},
  author={Bell, AR},
  journal={Monthly Notices of the Royal Astronomical Society},
  volume={182},
  number={2},
  pages={147--156},
  year={1978},
  publisher={Oxford University Press Oxford, UK}
}

@article{reed1995three,
  title={The three-dimensional structure of the Cassiopeia A supernova remnant. I. The spherical shell},
  author={Reed, Jeri E and Hester, J Jeff and Fabian, AC and Winkler, PF},
  journal={Astrophysical Journal v. 440, p. 706},
  volume={440},
  pages={706},
  year={1995}
}

@article{baars1977absolute,
  title={The absolute spectrum of CAS A-an accurate flux density scale and a set of secondary calibrators},
  author={Baars, JWM and Genzel, R and Pauliny-Toth, IIK and Witzel, A},
  journal={Astronomy and Astrophysics, vol. 61, no. 1, Oct. 1977, p. 99-106.},
  volume={61},
  pages={99--106},
  year={1977}
}

@article{krause2008cassiopeia,
  title={The Cassiopeia A supernova was of type IIb},
  author={Krause, Oliver and Birkmann, Stephan M and Usuda, Tomonori and Hattori, Takashi and Goto, Miwa and Rieke, George H and Misselt, Karl A},
  journal={Science},
  volume={320},
  number={5880},
  pages={1195--1197},
  year={2008},
  publisher={American Association for the Advancement of Science}
}

@article{bell1975new,
  title={New radio map of Cassiopeia A at 5 GHz},
  author={Bell, AR and Gull, SF and Kenderdine, S},
  journal={Nature},
  volume={257},
  number={5526},
  pages={463--465},
  year={1975},
  publisher={Nature Publishing Group UK London}
}

@article{delaney2014density,
  title={The density and mass of unshocked ejecta in Cassiopeia a through low frequency radio absorption},
  author={DeLaney, Tracey and Kassim, Namir E and Rudnick, Lawrence and Perley, RA},
  journal={The Astrophysical Journal},
  volume={785},
  number={1},
  pages={7},
  year={2014},
  publisher={IOP Publishing}
}

@article{maeda2009suzaku,
  title={Suzaku X-ray imaging and spectroscopy of Cassiopeia A},
  author={Maeda, Yoshitomo and Uchiyama, Yasunobu and Bamba, Aya and Kosugi, Hiroko and Tsunemi, Hiroshi and Helder, Eveline A and Vink, Jacco and Kodaka, Natsuki and Terada, Yukikatsu and Fukazawa, Yasushi and others},
  journal={Publications of the Astronomical Society of Japan},
  volume={61},
  number={6},
  pages={1217--1228},
  year={2009},
  publisher={Oxford University Press Oxford, UK}
}

@article{allen1997evidence,
  title={Evidence of X-ray synchrotron emission from electrons accelerated to 40 TeV in the supernova remnant Cassiopeia A},
  author={Allen, GE and Keohane, JW and Gotthelf, EV and Petre, R and Jahoda, K and Rothschild, RE and Lingenfelter, RE and Heindl, WA and Marsden, D and Gruber, DE and others},
  journal={The Astrophysical Journal},
  volume={487},
  number={1},
  pages={L97},
  year={1997},
  publisher={IOP Publishing}
}

@article{wang2016hard,
  title={Hard X-ray emissions from Cassiopeia A observed by INTEGRAL},
  author={Wang, Wei and Li, Zhuo},
  journal={The Astrophysical Journal},
  volume={825},
  number={2},
  pages={102},
  year={2016},
  publisher={IOP Publishing}
}

@article{aharonian2001evidence,
  title={Evidence for TeV gamma ray emission from Cassiopeia A},
  author={Aharonian, F and Akhperjanian, A and Barrio, J and Bernl{\"o}hr, K and B{\"o}rst, H and Bojahr, H and Bolz, O and Contreras, J and Cortina, J and Denninghoff, S and others},
  journal={Astronomy \& Astrophysics},
  volume={370},
  number={1},
  pages={112--120},
  year={2001},
  publisher={EDP Sciences}
}

@article{albert2007observation,
  title={Observation of VHE $\gamma$-rays from Cassiopeia A with the MAGIC telescope},
  author={Albert, J and Aliu, E and Anderhub, H and Antoranz, P and Armada, A and Baixeras, C and Barrio, JA and Bartko, H and Bastieri, Denis and Becker, JK and others},
  journal={Astronomy \& Astrophysics},
  volume={474},
  number={3},
  pages={937--940},
  year={2007},
  publisher={EDP Sciences}
}

@article{abdo2010fermi,
  title={Fermi-LAT discovery of GeV gamma-ray emission from the young supernova remnant Cassiopeia A},
  author={Abdo, AA and Ackermann, Markus and Ajello, Marco and Allafort, A and Baldini, Luca and Ballet, Jean and Barbiellini, Guido and Baring, MG and Bastieri, Denis and Baughman, BM and others},
  journal={The Astrophysical journal letters},
  volume={710},
  number={1},
  pages={L92},
  year={2010},
  publisher={IOP Publishing}
}

@article{ahnen2017cut,
  title={A cut-off in the TeV gamma-ray spectrum of the SNR Cassiopeia A},
  author={Ahnen, Max L and Ansoldi, S and Antonelli, LA and Arcaro, C and Babi{\'c}, A and Banerjee, B and Bangale, P and Barres de Almeida, U and Barrio, JA and Becerra Gonz{\'a}lez, J and others},
  journal={Monthly Notices of the Royal Astronomical Society},
  volume={472},
  number={3},
  pages={2956--2962},
  year={2017},
  publisher={Oxford University Press}
}

@article{cao2025broadband,
  title={Broadband $\gamma $-ray spectrum of supernova remnant Cassiopeia A},
  author={Cao, Zhen and Aharonian, F and Bai, YX and Bao, YW and Bastieri, D and Bi, XJ and Bi, YJ and Bian, W and Bukevich, AV and Cai, CM and others},
  journal={arXiv preprint arXiv:2502.04848},
  year={2025}
}

@article{hwang2004million,
  title={A million second Chandra view of Cassiopeia A},
  author={Hwang, Una and Laming, J Martin and Badenes, Carles and Berendse, Fred and Blondin, John and Cioffi, Denis and DeLaney, Tracey and Dewey, Daniel and Fesen, Robert and Flanagan, Kathryn A and others},
  journal={The Astrophysical Journal},
  volume={615},
  number={2},
  pages={L117},
  year={2004},
  publisher={IOP Publishing}
}

@article{fesen2006expansion,
  title={The expansion asymmetry and age of the Cassiopeia A supernova remnant},
  author={Fesen, Robert A and Hammell, Molly C and Morse, Jon and Chevalier, Roger A and Borkowski, Kazimierz J and Dopita, Michael A and Gerardy, Christopher L and Lawrence, Stephen S and Raymond, John C and Van Den Bergh, Sidney},
  journal={The Astrophysical Journal},
  volume={645},
  number={1},
  pages={283},
  year={2006},
  publisher={IOP Publishing}
}

@article{koldobskiy2021energy,
  title={Energy spectra of secondaries in proton-proton interactions},
  author={Koldobskiy, S and Kachelrie{\ss}, M and Lskavyan, A and Neronov, A and Ostapchenko, S and Semikoz, DV},
  journal={Physical Review D},
  volume={104},
  number={12},
  pages={123027},
  year={2021},
  publisher={APS}
}

@ARTICLE{naima,
   author = {{Zabalza}, V.},
    title = {naima: a Python package for inference of relativistic particle
             energy distributions from observed nonthermal spectra},
     year = 2015,
  journal = {Proc.~of International Cosmic Ray Conference 2015},
    pages = "922",
   eprint = {1509.03319},
   adsurl = {http://adsabs.harvard.edu/abs/2015arXiv150903319Z},
}

@article{zhan2022asymmetrical,
  title={An asymmetrical model for high-energy radiation of Cassiopeia A},
  author={Zhan, Shihong and Wang, Wei and Mou, Guobin and Li, Zhuo},
  journal={Monthly Notices of the Royal Astronomical Society},
  volume={513},
  number={2},
  pages={2471--2477},
  year={2022},
  publisher={Oxford University Press}
}

@article{drury1983introduction,
  title={An introduction to the theory of diffusive shock acceleration of energetic particles in tenuous plasmas},
  author={Drury, L O'C},
  journal={Reports on Progress in Physics},
  volume={46},
  number={8},
  pages={973},
  year={1983},
  publisher={IOP Publishing}
}

@article{vinyaikin2014frequency,
  title={Frequency dependence of the evolution of the radio emission of the supernova remnant Cas A},
  author={Vinyaikin, EN},
  journal={Astronomy Reports},
  volume={58},
  number={9},
  pages={626--639},
  year={2014},
  publisher={Springer}
}

@article{lee2014x,
  title={X-ray observation of the shocked red supergiant wind of Cassiopeia A},
  author={Lee, Jae-Joon and Park, Sangwook and Hughes, John P and Slane, Patrick O},
  journal={The Astrophysical Journal},
  volume={789},
  number={1},
  pages={7},
  year={2014},
  publisher={IOP Publishing}
}

@article{atoyan2000gamma,
       author = {{Atoyan}, A.~M. and {Aharonian}, F.~A. and {Tuffs}, R.~J. and {V{\"o}lk}, H.~J.},
        title = "{On the gamma-ray fluxes expected from Cassiopeia A}",
      journal = {\aap},
         year = 2000,
        month = mar,
       volume = {355},
        pages = {211-220},
          doi = {10.48550/arXiv.astro-ph/0001186},
archivePrefix = {arXiv},
       eprint = {astro-ph/0001186},
 primaryClass = {astro-ph},
       adsurl = {https://ui.adsabs.harvard.edu/abs/2000A&A...355..211A}
}

@article{mezger1986maps,
  title={Maps of Cassiopeia A and the Crab Nebula at lambda 1.2 MM},
  author={Mezger, PG and Tuffs, RJ and Chini, R and Kreysa, E and Gemuend, H-P},
  journal={Astronomy and Astrophysics (ISSN 0004-6361), vol. 167, no. 1, Oct. 1986, p. 145-150.},
  volume={167},
  pages={145--150},
  year={1986}
}

@article{de2016dust,
  title={The dust mass in Cassiopeia A from a spatially resolved Herschel analysis},
  author={De Looze, Ilse and Barlow, Mike J and Swinyard, Bruce M and Rho, Jeonghee and Gomez, Haley L and Matsuura, M and Wesson, Roger},
  journal={Monthly Notices of the Royal Astronomical Society},
  pages={stw2837},
  year={2016},
  publisher={Oxford University Press}
}

@article{arnaud2016planck,
  title={Planck intermediate results-XXXI. Microwave survey of Galactic supernova remnants},
  author={Arnaud, Monique and Ashdown, Mark and Atrio-Barandela, F and Aumont, J and Baccigalupi, C and Banday, AJ and Barreiro, RB and Battaner, E and Benabed, K and Benoit-L{\'e}vy, A and others},
  journal={Astronomy \& Astrophysics},
  volume={586},
  pages={A134},
  year={2016},
  publisher={EDP sciences}
}

@article{de2020cassiopeia,
  title={Cassiopeia A, Cygnus A, Taurus A, and Virgo A at ultra-low radio frequencies},
  author={De Gasperin, F and Vink, J and McKean, JP and Asgekar, A and Avruch, I and Bentum, MJ and Blaauw, R and Bonafede, A and Broderick, JW and Br{\"u}ggen, M and others},
  journal={Astronomy \& Astrophysics},
  volume={635},
  pages={A150},
  year={2020},
  publisher={EDP Sciences}
}

@article{stanislavsky2023free,
  title={Free--free absorption parameters of Cassiopeia A from low-frequency interferometric observations},
  author={Stanislavsky, Lev A and Bubnov, Igor N and Konovalenko, Aleksander A and Stanislavsky, Aleksander A and Yerin, Serge N},
  journal={Astronomy \& Astrophysics},
  volume={670},
  pages={A157},
  year={2023},
  publisher={EDP Sciences}
}

@article{arias2018low,
  title={Low-frequency radio absorption in Cassiopeia A},
  author={Arias, Maria and Vink, J and de Gasperin, Francesco and Salas, P and Oonk, JBR and Van Weeren, RJ and Van Amesfoort, AS and Anderson, James and Beck, Rainer and Bell, ME and others},
  journal={Astronomy \& Astrophysics},
  volume={612},
  pages={A110},
  year={2018},
  publisher={EDP sciences}
}

@article{humensky2008veritas,
  title={VERITAS studies of the supernova remnants Cas A and IC 443},
  author={Humensky, Thomas Brian},
  journal={arXiv preprint arXiv:0810.0799},
  year={2008}
}

@article{abeysekara2020evidence,
  title={Evidence for proton acceleration up to TeV energies based on VERITAS and Fermi-LAT observations of the Cas A SNR},
  author={Abeysekara, AU and Archer, A and Benbow, W and Bird, R and Brose, R and Buchovecky, M and Buckley, JH and Chromey, AJ and Cui, W and Daniel, MK and others},
  journal={The Astrophysical Journal},
  volume={894},
  number={1},
  pages={51},
  year={2020},
  publisher={IOP Publishing}
}

@article{guberman2017deep,
  title={Deep observations of Cas A with MAGIC indicate it is no PeVatron},
  author={Guberman, D and Cortina, J and Wilhelmi, E and Galindo, D and Moralejo, A},
  journal={arXiv preprint arXiv:1709.00280},
  year={2017}
}

@article{cao2024does,
  title={Does or did the supernova remnant Cassiopeia A operate as a PeVatron?},
  author={Cao, Zhen and Aharonian, F and An, Q and Bai, YX and Bao, YW and Bastieri, D and Bi, XJ and Bi, YJ and Cai, JT and Cao, Q and others},
  journal={The Astrophysical Journal Letters},
  volume={961},
  number={2},
  pages={L43},
  year={2024},
  publisher={IOP Publishing}
}

@article{zhang2019supernova,
  title={Is Supernova Remnant Cassiopeia A a PeVatron?},
  author={Zhang, Xiao and Liu, Siming},
  journal={The Astrophysical Journal},
  volume={874},
  number={1},
  pages={98},
  year={2019},
  publisher={IOP Publishing}
}

@ARTICLE{atoyan2000energy,
       author = {{Atoyan}, A.~M. and {Tuffs}, R.~J. and {Aharonian}, F.~A. and {V{\"o}lk}, H.~J.},
        title = "{On energy-dependent propagation effects and acceleration sites of relativistic electrons in Cassiopeia A}",
      journal = {\aap},
         year = 2000,
        month = feb,
       volume = {354},
        pages = {915-930},
          doi = {10.48550/arXiv.astro-ph/0001187},
archivePrefix = {arXiv},
       eprint = {astro-ph/0001187},
 primaryClass = {astro-ph},
       adsurl = {https://ui.adsabs.harvard.edu/abs/2000A&A...354..915A}
}

\end{document}